\documentclass[12pt]{article} %[handout] %[colordvi]              
\usepackage[german,english]{babel}

\usepackage{amscd,amsmath}
\usepackage {amsfonts,apalike}
\usepackage{epsfig,graphicx,graphics}

\usepackage{cancel}

\usepackage{undertilde}
\usepackage[normalem]{ulem} % perhaps strikethrought \sout{}

\usepackage{epsfig,graphicx,graphics}

\usepackage{cancel}
\DeclareGraphicsExtensions{.pdf,.png,.jpg}

\UseRawInputEncoding

\title{Peter Bergmann on Observables in Hamiltonian General Relativity: A Historical-Critical Investigation\\ \emph{Studies in History and Philosophy of Science} {\bf 95} (2022) pp. 1-27, doi: 10.1016/j.shpsa.2022.06.012. }

 \author{J. Brian Pitts \\Faculty of Philosophy,  University of Cambridge,  \\ Programme in Philosophy,  University of Lincoln, and \\ Department of Philosophy, University of South Carolina  \\    Funded by the National Science Foundation (USA) \#1734402} % 

 \date{ 7 June 2022} %\today} 

\begin{document}

\maketitle
% Creates title page of slide show using above information

\begin{abstract}

The problem of observables and their supposed lack of change has been significant in Hamiltonian quantum gravity since the 1950s.  This paper considers the unrecognized variety of ideas about observables in the thought of constrained Hamiltonian dynamics co-founder Peter Bergmann, who trained many students at Syracuse and invented observables.  Whereas initially Bergmann required a constrained Hamiltonian formalism to be mathematically equivalent to the Lagrangian, in 1953 Bergmann and Schiller introduced a novel postulate, motivated by facilitating quantum gravity.  This postulate held that observables were \emph{invariant} under transformations generated by \emph{each individual} first-class constraint.  While modern works rely on Bergmann's authority and sometimes speak of ``Bergmann observables,'' he had much to say  about observables, generally  interesting and plausible but not all mutually  consistent and much of it neglected.

On occasion he required observables to be locally defined (not changeless and global); at times he wanted observables to be independent of the Hamiltonian formalism (implicitly contrary to a definition involving separate first-class constraints). But typically he took observables to have vanishing Poisson bracket with each first-class constraint and took this result to be justified by the example of electrodynamics.  He expected observables to be analogous to the transverse true degrees of freedom of electromagnetism.  Given these premises, there is no coherent concept of observables which he reliably endorsed, much less established. 

 A revised definition of observables that satisfies the requirement that equivalent theories should have equivalent observables using the Rosenfeld-Anderson-Bergmann-Castellani gauge generator $G$, a tuned sum of first-class constraints that changes the canonical action $\int dt(p\dot{q}-H)$ by a boundary term. Bootstrapping from theory formulations with no first-class constraints, one finds that the ``external'' coordinate gauge symmetry of GR calls for covariance (a transformation rule and hence a $4$-dimensional Lie derivative for the Poisson bracket), not invariance ($0$ Poisson bracket), under $G$ (not each first-class constraint separately).

\end{abstract}

%%%%%%%%%%%%%%%%%%%%%%%%%%%%%%%%%%%%%%%%%%%%%%%%%%%%%%%%%%%%%%%

\section{Introduction}

The problem of observables in Hamiltonian General Relativity (GR) finds its origins in the many works of Peter G. Bergmann and his various collaborators (mostly his students) starting in the 1950s.  Hamiltonian methods are the simplest way to quantize a theory, hence the term ``canonical'' quantization.  Unfortunately, the method requires reinvention for Maxwell's electromagnetism and Einstein's General Relativity (GR) because these theories do not permit the traditional Legendre transformation from velocities to momenta.\footnote{At any rate some modification is needed; Fermi's trick was another possibility.}   This impossibility characterizes theories with gauge freedom (such as those examples), theories with `broken' gauge freedom (such as Proca's massive electromagnetism or massive gravity, a topic of renewed attention in the last decade or so), and spinor fields such as describe electrons (for which the Lagrangian is linear in velocities) \cite{SpinorChange}.  Hence there arose a need to reinvent Hamiltonian dynamics as ``constrained Hamiltonian dynamics'' (\emph{e.g.}, \cite{RosenfeldQG,BergmannNonlinear,DiracCanadian,Sundermeyer}).  Whereas naively in classical mechanics one takes all mathematical quantities to be observable, gauge freedom introduces a  new question of what is observable once some distinctions are held to make no physical difference to the state or at least to the history of the world. A possibly related and potentially confusing further question is the issue of quantum observables. One should not assume that observables and quantum mechanics and observables in General Relativity or other classical gauge field theories are analogous \cite[p. 105]{Kiefer3rd}.

Early work in canonical quantum gravity recently has received considerable attention including a reprint volume and a monograph, as well as a translation and commentary on Rosenfeld's groundbreaking work---work that was not continued or even remembered for most of two decades 
\cite{RosenfeldQG,SalisburySundermeyerRosenfeldQG,SalisburyRosenfeldMed,RicklesBlumWeissCanonical,SalisburySyracuse1949to1962,RicklesQGandHPS,BlumSalisburyRickles,RicklesDeepMistQG}.   
This paper has some similarities to Salisbury's account, especially regarding the loss of aspiration to exhibit $4$-dimensional coordinate symmetry in canonical or canonically quantized General Relativity \cite{SalisburySyracuse1949to1962,SalisburyObservablesHJ}. Salisbury explains how Bergmann's enthusiastic, perhaps excessive, admiration for Dirac's contributions to Hamiltonian GR helped Bergmann to discard 40\% of the space-time metric and the associated  conjugate momenta as conjugate variables and hence to lose hold of $4$-dimensional space-time symmetry to a considerable degree.   
One could likely find an example of ``Kuhn loss,'' the disappearance of knowledge under change of paradigm (\emph{c.f.} \cite{HoyningenHueneKuhnLoss,VerronenKuhnLoss}) somewhere in the loss of space-time covariance from the 1950s onward. 

  This paper is, however, more narrowly focussed on observables (not on Hamilton-Jacobi theory) and also  more skeptical of the coherence of Bergmann's package of ideas than is Salisbury's.  This skepticism is motivated by the normative considerations at the beginning  of the paper, which briefly review ideas developed elsewhere \cite{FirstClassNotGaugeEM,GRChangeNoKilling,ObservablesEquivalentCQG,ObservablesLSEFoP,ObservablesEinsteinMaxwellFoP}, 
 with the primarily historical Bergmann-focussed work following  with interspersed normative comments justified by the introductory material.  One can view this paper as an effort to explain how large portions of the physics and philosophy communities arrived at paradoxical conclusions about the apparent absence of time and change in General Relativity.  

 The question of observables is not merely philosophical, because some programs in quantum gravity have continued to make essential use of the concept of observables \cite{ThiemannReducedQuantization,GiddingsMarolfHartleObservables,DittrichPartialConstrained}.  Hence investigating the foundations of Bergmann's widely received ideas might shed useful light on quantum gravity.  It might turn out, for example, that such projects pertain not to observables, but to another  important notion, perhaps such as true degrees of freedom, which clearly were part of what Bergmann long sought in his work on observables.    
 Chataignier explains how to find equivalence among definitions of observables that appear to disagree: 
\begin{quote} 
Therefore, the views regarding the definition of observables discussed above (observables
as gauge-fixed quantities; relational observables; observables as quantities that
transform ``covariantly'') are, in fact, equivalent. \cite{ChataignierDissertation} 
\end{quote} 
Hence revising one's opinions about observables might have few or no technical ramifications for quantum gravity, but quantum gravity would have less of a paradoxical air.  Elsewhere I have noted parallels with Rovelli's work on partial observables \cite{RovelliObservable,RovelliPartialObservables,GRChangeNoKilling,SpinorChange}, as well as how Kuch\v{a}r's \cite{KucharCanonical93} dissatisfaction with traditional definitions helped to motivate my work.  One should also mention Smolin's critique of the usual definition of observables and preference that observables be related to observations \cite{SmolinPresent}.

%%%%%%%%%%%%%%%%%%%%%%%%%%%%%%%%%%%%%%%%%%%%%%%%%%

Starting in the late 1940s, Bergmann, a former assistant of Einstein, (re)invented constrained Hamiltonian dynamics.  Dirac did parallel work around the same time, but I will be interested in Dirac's work only occasionally and in the appendix. Rosenfeld's earlier pioneering work is receiving increased appreciation and even translation recently \cite{RosenfeldQG,SalisburySundermeyerRosenfeldQG}.  Bergmann founded a key GR and quantum gravity group at Syracuse, and that at a time when General Relativity was typically studied by mathematicians if it was studied at all; this was prior to the Renaissance of General Relativity \cite{EisenstaedtEtiage,Eisenstaedt,SchutzGR,BlumLalliRennGRRenaissance,BlumGiuliniLalliRennEPJHintro,BlumLalliRennIsis,BlumLalliRennGRRenaissanceEinsteinStudies}.\footnote{The dominance of two figures in the invention of constrained dynamics and the relative recency of peer review might be worth considering regarding quality control and how insufficiently tested ideas could become widely received.}  
The problem of observables still afflicts canonical quantum gravity. There are rival definitions, at least some of which are assigned key technical roles in certain quantization programs, but there seems to be a default status assigned to a definition that Bergmann at times used---even if Dirac's name is attached to it (``Dirac observables'')! This definition has the peculiar consequence of implying that observables are constants of the motion and require integration over the whole universe \cite{TorreObservable}, rather than being local fields analogous to the electromagnetic field strength $F_{\mu\nu}$ in electromagnetism or the Ricci curvature tensor $R_{\mu\nu}$ in GR as one might have expected on Lagrangian grounds.  
  It is therefore important to understand where definitions came from and what their justification might be. This process might even help to diagnose and cure conceptual-technical problems faced in ongoing canonical quantization programs.  At a minimum, it will help to show whether the problem of missing change exists already at the classical level (as has been argued \cite{EarmanMcTaggart}) or rather arises at the quantum level.  I have elsewhere argued  that the problem is purely a verbal one at the classical level due to multiple unjustified and indeed erroneous partial definitions \cite{GRChangeNoKilling,ObservablesEquivalentCQG,ObservablesLSEFoP,ObservablesEinsteinMaxwellFoP}. While one can define a new word however one likes, one cannot do the same to a familiar word and retain assurance that its old associations still hold; one is likely to introduce equivocation.  A definition can be wrong in  the sense of implying contradictions with  more certain or even unrevisable claims.

This paper attempts to survey nearly everything that Bergmann wrote that addresses the subject of observables in General Relativity, while also making a normative evaluation in terms of  mathematical (not verbal-stipulative) standards of equivalence to the Lagrangian and empirical equivalence.  It will emerge that Bergmann had a generous list of desiderata that observables ought to satisfy, each of which is individually plausible and most  of which are co-instantiated in electromagnetism, but which simply do not all fit together in the case of General Relativity. There simply isn't anything that has all of the features in question.  While one can certainly postulate long lists of desired properties for observables, nothing ensures that the world, or more to the point General Relativity, actually contains anything satisfying such a description, or contains nearly enough of them with the expected roles. This non-existence invalidates many claims about what features  observables supposedly  have; merely including properties in a definition does not imply that one is thereby talking about any entities that have those properties, because nothing has all those properties.  By refining our criteria to include only the best justified and mutually compatible criteria, we can hope to have a reasonable concept that really is exemplified and that will serve at least some of the purposes that observables have been intended to serve, such as being related to observations (a requirement that Bergmann often endorsed). The multiplicity of notions in Bergmann's thought renders it questionable to continue using the phrase ``Bergmann observables,'' a phrase sometimes attached to quantities that are supposed to be predictable in the sense of a Cauchy problem.  This phrase is questionable both on historiographic grounds for misrepresenting Bergmann's views and on normative grounds for picking out a concept that is not very interesting---either uninstantiated or at best rather sparse, weakly motivated, and unable to fulfill Bergmann's intentions.  The phrase ``Dirac observables,'' while perhaps justifiable historiographically in the sense of being what Dirac would have said about observables if pressed (as well as part of what Bergmann said!), suffers normatively from two different objections, one involving the separate use of first-class constraints and one involving the neglect of the internal \emph{vs.} external gauge symmetry distinction or something in that vicinity distinguishing electromagnetism from General Relativity.  An appendix discusses Dirac's work briefly. 

This normatively tinged history  also yields at least a partial solution to the problem of missing change (an aspect of the ``problem of time''), in that observables, suitably redefined, are spatiotemporally varying local fields, basically (for transformations near the identity at least), scalars, vectors, tensors, \emph{etc.}, or more generally geometric objects (in the classical sense of a set of components relative to a local coordinate system along with a transformation rule \cite{Nijenhuis}) that are also invariant under internal (\emph{e.g.}, electromagnetic or Yang-Mills) gauge transformations.  Hence observables should be invariant under internal gauge transformations and covariant under coordinate transformations, much as one would expect on non-Hamiltonian grounds \cite{GRChangeNoKilling,ObservablesEquivalentCQG,ObservablesLSEFoP,ObservablesEinsteinMaxwellFoP}. Previous work generally has not applied the internal \emph{vs.} external distinction, with the partial and fleeting exception of Bergmann's contemplation of splitting coordinate transformations into what he envisaged as a (internal?) gauge transformation requiring invariance and a (external?) Lorentz transformation requiring covariance \cite{BergmannLorentz,BergmannRadiationObservables}.    Such a definition of observables appears to give reasonable answers in all known cases with the possible exception of locally supersymmetric theories such as  supergravity and superstring theory, for which the mixing of internal and external gauge symmetries \cite{vanNReports} poses a potential challenge to a definition that treats the two separately. But it is not obvious  what a translation by an \emph{anticommuting} distance means or how one makes observations in supergravity \cite{IsenbergSUGRAClassical}, so this challenge is likely to disappear on closer inspection. Hence reflective equilibrium between generalities and examples is achieved or at least approached.  

As a matter of scope, this paper does not attempt to study systematically  what Bergmann's early students  or collaborators (such as Arthur Komar) produced on other occasions independently.  (In only some cases their dissertations supervised by Bergmann have been used.)  Of such works, some of James Anderson's are highly relevant, such as (\cite{AndersonChange}).  Anderson's talk about Lie derivatives and all possible vector fields is in the vicinity of the recent   no time-like Killing vector condition \cite{GRChangeNoKilling} as the criterion for change in vacuum GR.  But Anderson by then was too committed to the idea that a first-class constraint generates a gauge transformation \cite{AndersonObservables}  (by itself rather than as part of the team $G$) and that the correct Poisson bracket is a vanishing one (not a Lie derivative) to take the non-existence of a time-like Killing vector as the standard for change.  (This criterion is, however, widely used in the GR exact solutions literature \cite{ExactSolns2} to distinguish ``stationary'' (unchanging) and ``non-stationary'' (changing) solutions). Anderson somehow retained some of his earlier use of the gauge generator $G$ alongside individual use of first-class constraints \cite{AndersonQGR,AndersonCoordinate}.  One can find Anderson claiming that  $\mathcal{H}_L$ and $\mathcal{H}^r$ (which came to be called the Hamiltonian constraint and the momentum constraint, the former denoted by $\mathcal{H}_0$ below, the latter here in contravariant form but later in covariant form as $\mathcal{H}_i$) generate coordinate transformations and then citing his 1951 paper with Bergmann that apparently contained no such ideas and  (re)invented the gauge generator $G$, a tuned sum of first-class constraints \cite{AndersonCoordinate,AndersonBergmann}. Manifestly the constraints $\mathcal{H}_L$ and $\mathcal{H}^r$ do nothing to the time-space parts of the metric (\emph{e.g.}, the lapse function and shift vector) and hence do not implement spatial-\emph{temporal} coordinate transformations or even spatial coordinate transformations on the space-time metric.  Anderson evidently intended a strong reading of the claim  ``that the degrees of freedom associated with $g_{0\mu}$ disappear from the Hamiltonian formalism.'' Exhaustive study of Bergmann's early students' works might flesh out somewhat the already luxuriant collection of ideas about observables in Bergmann's work, but would not uncover a coherent view that incorporates all of Bergmann's ideas.  J. Goldberg's historical review regarded the conventional Bergmann wisdom as true and important \cite{GoldbergSyracuse}. 
Salisbury has ably surveyed Bergmann's contributions to constrained Hamiltonian dynamics, primarily  focusing on the mathematically formalism \cite{SalisburyBergmann}. He briefly discusses the idea of observables, quoting some correspondence in  1959 between Bergmann and Dirac about the surprising apparent constancy over time of various quantities.  By contrast this paper makes little effort to pursue unpublished sources, worthy as such a task might be, because the published record is insufficiently studied, especially from my normative perspective.  It is hoped that this paper will fill a substantial portion of that gap.

The more normative parts of this work in some respects parallel the very sophisticated work of Salisbury, Renn and Sundermeyer \cite{SalisburyRennSundermeyerHamiltonJacobi}, such as in their criticism of the conventional wisdom about a phase space using only the spatial metric and its canonical momenta and in their emphasis on retaining the lapse and shift as canonical variables.  Those authors employ curvature scalars as intrinsic coordinates, expect invariance (not covariance) under the gauge generator $G$, and  employ active diffeomorphisms, however. A few passages illustrate themes that will also appear below in discussions of intrinsic coordinates. ``Yet, perhaps paradoxically, the liberty in selecting intrinsic coordinates is precisely as broad as is the original diffeomorphism freedom.''  ``For every choice of coordinate parameters in general relativity there corresponds a choice of intrinsic coordinates.''  I believe that a formalism that uses merely passive coordinate transformations is equivalent, but leaner in avoiding introducing an extra copy (or two) of the coordinate freedom \cite{ObservablesEquivalentCQG}. If one considers behavior relative to different choices of intrinsic coordinates, one is likely to need a (coordinate) transformation rule and hence a form of covariance as opposed to invariance.

%%%%%%%%%%%%%%%%%%%%%%%%%%%%%%%%%%%%%%%%%%%%%%%%%%%%%%%%%%%%%%%%%%%%%%%%%%%%%%%%%%%%%%%%%%%

\section{Rosenfeld-Dirac-Bergmann Constrained Hamiltonian Dynamics}

\subsection{Overview} 

To make this paper sufficiently self-contained, a brief reminder of Rosenfeld-Dirac-Bergmann constrained Hamiltonian dynamics is appropriate. 
 Modern  fundamental physics is primarily comprised of gauge fields  (broadly construed in the sense of having arbitrary functions of space-time in the solutions of the field equations), perhaps what one might call broken gauge fields (such as a would-be gauge field plus a mass term), and certainly spinors fields (used to represent fermions after quantization). While constrained Hamiltonian dynamics is sociologically a slightly exotic topic,  most of fundamental physics involves theories for which the traditional Legendre transformation to the Hamiltonian is impossible, making ordinary Hamiltonian physics the oddity.  For the familiar case of Maxwell's electromagnetism,  $\mathcal{L}=-\frac{1}{4}F_{\mu\nu}F^{\mu\nu}$ and  $F_{\mu\nu} =_{df} \frac{ \partial A_{\nu}}{\partial x^{\mu}   } - \frac{ \partial A_{\mu}}{\partial x^{\nu}   }.$
While one can always define canonical momenta in the usual way, one is not guaranteed that the result will fulfill one's hopes to replace $\dot{q}$ with $p$.  In this case changing from $\dot{q}^{\mu}$ to $p_{\mu}$ fails because  $p_{\mu}=_{df} \frac{\partial \mathcal{L} }{\partial \dot{q}_{\mu} } $ is not fully soluble for $\dot{q}^{\mu}$. 
Thus one has some   ``primary constraints,'' which  fix  some of the  $p_{\mu},$ such as to $0$ in simple cases, or more generally to a function of the $q$'s and other $p$'s.
Hence for Maxwell's massless (or Proca's massive) electromagnetism, $p^0(x)=_{df} \frac{\partial \mathcal{L} }{\partial A_0,_0 }=0. $    For GR  (or massive GR),  everything is analogous but much harder.  A Lagrangian (or its kinetic term in the massive case) is  $\mathcal{L}= \frac{1}{16 \pi G}\sqrt{-g}R(g,\partial g, \partial^2 g) + \mathcal{E}^{\mu},_{\mu}$  and the gauge freedom is passive coordinate transformations  $g_{\mu\nu} \rightarrow g_{\mu\nu} + \pounds_{\xi}g_{\mu\nu}$.
Given a wise choice of  $\mathcal{E}^{\mu},_{\mu}$  \cite{DiracHamGR,AndersonPrimary,BergmannKomarLeeHJ} as well as a wise choice of field variables as the metric $g_{\mu\nu}$ (or something sufficiently close, such as the lapse function $N$, shift vector\footnote{I use a hybrid of two common notations, $N$ for the lapse (thus leaving $\alpha$ free for the weight $-1$ slicing density \cite{JantzenYork}) and $\beta^i$ for the shift, thus avoiding any potential confusion between components of the shift and powers of the lapse that could arise from the expression $N^i,$ if not in GR, then in massive variants of GR or other alternative theories.} 
 $\beta^i,$ and $3$-metric $g_{ij}$)\footnote{By contrast, for these purposes some otherwise attractive but unwise choices of field variables include $g^{\mu\nu}$, $\mathfrak{g}^{\mu\nu}= \sqrt{-g}g^{\mu\nu},$  $\mathfrak{g}_{\mu\nu} = g_{\mu\nu}/\sqrt{-g},$ \emph{etc.}}, the primary constraints take the trivial form of annihilating certain momenta:  $p(x)=_{df} \frac{\partial \mathcal{L} }{\partial {N},_{0}(x) }=0$,  $p_{i}(x)=_{df} \frac{\partial \mathcal{L} }{\partial {\beta}^i,_{0} }=0$.  (Henceforth spatial arguments will be surprised where possible.) Thus 40\% of the Legendre transformation cannot be done at all, while the remaining 60\% proceeds  as usual, given a suitable Lagrangian and a suitable choice of field variables (neither of which is quite unique, but the wise options are few).  This material is now so standard \cite{MTW,Sundermeyer,Wald}, sometimes omitting the primary constraints altogether, that one might fail to  appreciate the challenges of Hamiltonian GR faced in the early 1950s in the nontrivial primary constraints \cite{PiraniSchild,PiraniSchildSkinner,DeWittSpinor,BelinfanteCK}. Recently some authors dispensed with  such simplifications \cite{Kiriushcheva} in favor of heroic calculations.

If some canonical momenta are set to $0$ (or otherwise fixed) by the degenerate Legendre transformation, then achieving consistency with Hamiltonian time evolution for such momenta requires that such primary constraints be preserved over time:  the momenta have to \emph{stay} $0$ (or some such, constrained in terms of the $q$'s and other $p$'s) \cite{Sundermeyer}.  Making the Hamiltonian (of which a precise definition has not been given here) preserve primary  constraints in many cases gives ``secondary constraints,'' functions of perhaps $q,$ $p,$  $q,_i$, $p,_i$, and even $q,_{ij}$ that must be $0$. In electromagnetism and GR, the secondary constraints are phase space analogs of Gauss's law and the Gauss-Codazzi embedding equations for space into space-time, respectively, and so are both important and familiar.  In GR the secondary constraints $\mathcal{H}_0$ and $\mathcal{H}_i$ (met above with slightly different notation) are known (among other things) as the Hamiltonian constraint and the momentum constraint.   It is tempting to assign these constraints physical meanings in relation to gauge transformations.  But such meanings fail because the secondary constraints do not transform the scalar potential or the lapse and shift suitably \cite{FirstClassNotGaugeEM,GRChangeNoKilling}, as one sees immediately from brackets like $\{ \mathcal{H}_i(x), \beta^j(y) \} = 0$ and $\{ \mathcal{H}_0(x), N(y) \} =0.$    
One needs to insist on the dynamical preservation of the secondary constraints as well, iterating until the  algorithm terminates.
 For Maxwell's electromagnetism or GR, the algorithm terminates with the secondary constraints.  For some theories, for example, unimodular General Relativity and pure spin $2$ massive gravity, ``tertiary'' constraints arise \cite{UnruhUGR,Marzban,HassanRosen}.  For pure spin $2$ massive gravity, quarternary constraints also arise \cite{Marzban,HassanRosen}, though recent work on nonlinear massive pure spin $2$ gravity follows the shorthand of using  ``secondary'' as a catch-all to include tertiary and quarternary constraints.  
The taxonomy (due to Bergmann's school) of primary, secondary, tertiary, and quarternary, pertaining to the order in which the constraints arise from the formalism, is orthogonal to another classification (due to Dirac) with confusingly similar terminology, first-class \emph{vs}. second-class, which has quite different significance. In applications one sometimes has to take linear combinations of (\emph{e.g.}) primary and secondary constraints in order to find all the first-class constraints.

That other distinction, first-class \emph{vs.} second-class, requires more explicit attention to the notion of Poisson brackets.  In field theories one uses a field-theoretic  Poisson bracket: $$\{ \phi(x), \psi(y) \}=_{df} \int d^3z \sum_A \left( \frac{\delta \phi(x) }{\delta q^A(z) } \frac{\delta \psi(y)}{\delta p_A(z)}  - \frac{\delta \phi(x)}{\delta p_A(z)} \frac{\delta \psi(y)}{\delta q_A(z)} \right). $$   
It follows that   $\{ q^A(x), p_B(y) \} = \delta^A_B \delta(x,y),$ hence $0$ if not conjugate, as one would expect.  
One can now express the second distinction as between first-class and second-class constraints.  First-class constraints have $0$ Poisson brackets with all the constraints (possibly using the constraints themselves), while second-class constraints do not.  This distinction, though coarsely stated here, is important because  first-class constraints are associated (somehow!) with gauge freedom and second-class constraints are not, loosely speaking.  Several issues could be teased out here, such as the potential need to redefine the constraints to get the `right' number (as many as possible) to be first-class, as in Einstein-Proca theory (classically, massless gravitons and massive photons) \cite{EinsteinProca}, and more crucially for present purposes, the nature of the association between first-class constraints and gauge freedom, which is contested and crucial.

A watershed issue, one that will be confronted repeatedly in reviewing Bergmann's work on observables and gauge freedom, is whether first-class constraints generate gauge freedom \emph{as a team} \cite{RosenfeldQG,AndersonBergmann,CastellaniGaugeGenerator,FirstClassNotGaugeEM,SalisburySundermeyerRosenfeldQG}, the original view and the one indicated by requiring mathematical Hamiltonian-Lagrangian equivalence, or as individual constraints, which during the 1950s onward became the dominant view (especially in books).  This later view is supposed to be \emph{physically} equivalent to the Lagrangian---that is, equivalent for ``observables''---even while differing mathematically.  Clearly the status of any notion of observables used is an issue of the utmost importance, requiring not stipulation or tradition (as one actually finds typically), but mathematical proof.  
For our two stock examples,  Maxwell and Einstein (as well as the intermediate Yang-Mills case)  in the most common formalisms \cite{Sundermeyer}, all constraints are ``first class'':  they have Poisson brackets among themselves that are $0$ or proportional to the constraints themselves and hence have a value of $0$ on the constraint surface.  Given the connection between masslessness and gauge freedom for spins  $\geq 1$   \cite{FierzPauli}, it is therefore no surprise that all constraints in the usual formulations of these theories are first-class.

%%%%%%%%%%%%%%%%%%%%%%%%%%%%%%%

\subsection{First-Class Constraints and Gauge in More Detail}

  Exactly how do first-class constraints relate  to  gauge freedom?   On this question there are two standard answers.  There is the original, mathematically-derived answer that seeks in effect to make a Legendre transformation of the Lagrangian gauge freedom.  (Sometimes the result is more complicated due to the need to employ Hamilton's equations, as  happens in GR for changes of time coordinate \cite{FradkinVilkoviskyHLEquivalence} at least in the $3+1$ formulation.)  On this view, there is a  gauge generator $G$ \cite{RosenfeldQG,AndersonBergmann,CastellaniGaugeGenerator}, a tuned sum or team of first-class constraints that work together due to having related coefficients.  Thus for  Maxwell,  $ G=\int d^3x (p^i,_i \epsilon - p^0 \dot{\epsilon})$:  there are two first-class constraints (at each point), but they are combined using only one arbitrary function and its time derivative. The resulting gauge transformation is  $\delta A_{\mu} = \{ A_{\mu}, G \} = -\epsilon,_{\mu}.$  Despite being mathematically justified, this view gradually diminished in the 1950s and in some circles is quite unknown.  It has, however, been re-emphasized or indeed reinvented in the last 40-odd years by Mukunda, Castellani, Salisbury, Pons, Shepley, Sugano, Kimura, Sundermeyer and some others \cite{MukundaGaugeGenerator,Kamimura,CastellaniGaugeGenerator,SalisburySundermeyerEinstein,SuganoGaugeGenerator,GraciaPons,ShepleyEvolutionary,PonsSalisburyShepley,PonsReduce,ShepleyPonsSalisburyTurkish,PonsSalisbury,PonsSalisburySundermeyerFolklore,FirstClassNotGaugeEM}.  

  While no one can deny that the gauge generator $G$ performs as advertised, must one be so strict?  Since the mid-late 1950s a more common answer has been that  a first-class constraint \emph{by itself} generates a gauge transformation. On this view, for Maxwell one has  $ \delta A_{\mu}(x) = \{ A_{\mu}(x), \int d^3y [p^0(y) \xi(t,y)      + p^i,_i(y) \epsilon(t,y)] \}= \delta^0_{\mu} \xi  -\delta_{\mu}^i \partial_i \epsilon.$  It is not often noticed that this transformation changes  $\vec{E}$  \cite{FirstClassNotGaugeEM}.
%   Somehow incorrect answer flourishes despite correct one.
%  ``\ldots [V]ague relation between first class constraint transformations and local gauge transformations.''  \cite[p. 134]{Sundermeyer}. Working toward reinventing $G$.  
But perhaps such a result should be no surprise, because Anderson and Bergmann already pointed out that the preservation of Lagrangian constraint surface $\Sigma_l$ corresponds to transformations generated by $G$:  
\begin{quote}
Naturally, other forms of the hamiltonian density can be obtained by canonical transformations; but the arguments appearing in such new expressions will no longer have the significance of the original field variables $y_A$ and the momentum densities defined by Eq. (4.2) [which defines the canonical momenta as $\pi^A \equiv \frac{ \partial L}{\partial \dot{y}_A }$].  It follows in particular that transformations of the form (2.4) [``invariant'' transformations changing $\mathcal{L}$ by at most a divergence, such as electromagnetic gauge transformations or passive coordinate transformations in GR] will change the expression (4.9) [for the Hamiltonian density] at most by adding to it further linear combinations  of the primary constrains, i.e., by leading to new arbitrary functions $w^i.$  \cite[p. 1021]{AndersonBergmann} \end{quote}
The figure aims to illustrate their point: 

\vspace{-.25in} 
$$\begin{CD}
\mathcal{L}    @>?>> \mathcal{L}^{\prime}  \\
@VconstrainedVLegendreV @AinverseAconstrained\hspace{.05in}Legendre?A \\
\mathcal{H}  @>general\hspace{.05in}canonical\hspace{.05in}>violates\hspace{.05in}q_A\hspace{.05in}sense\hspace{.05in}or\hspace{.05in}4.2:\hspace{.05in}\pi^A=\frac{\partial\mathcal{L}}{\partial\dot{q}_A}> \mathcal{H}^{\prime}
\end{CD}$$

%%%%%%%%%%%%%%%%%%%%%%%%%%%%%%%%%%%%%%%%%%%
%%%%%%%%%%%%%%%%%%%%%%%%%%%%%%%%%%%%%%%%%%%%%%%%%%%

\subsection{What Is Real Change in GR?}

The idea that change is missing in Hamiltonian General Relativity, or at any rate from observables, dates back to the mid-1950s \cite{BergmannGoldberg}.  To gain perspective on how seriously to take this result, one can consider what one means by change, especially in a toy example in which explicit calculation is easy.  One possibility is a naively spatially   homogeneous theory, simply throwing away spatial dependence and consequently the shift vector $\beta^i$ and its conjugate momenta $p_i$  \cite{AshtekarBianchi,PonsReduce,GRChangeNoKilling}.  (There are issues regarding the number of degrees of freedom and remnants of spatial general covariance, but they are not important for my purposes.)  Then one sees that   $H = N \mathcal{H}_0 + \dot{N} p_0$ is reparametrization-invariant: simultaneity is fixed but can relabel the slices, bunching or spreading moments of label time (relative to proper time).  Projectability to phase space (eliminating unwanted velocities)  requires using not the time components but the  normal projection $\epsilon^{\perp}= N \xi^0$ as the primitive  descriptor of the transformation, having $0$ Poisson bracket with the canonical coordinates and momenta. 
The gauge generator is then $G = \epsilon^{\perp}  \mathcal{H}_0 + \dot{\epsilon}^{\perp} p$ \cite{CastellaniGaugeGenerator,GRChangeNoKilling}.   $G$ gives terms along the lines of Hamilton's equations multiplied by  $ \xi^0 $  (which are tolerable) and (importantly) the Lie derivative with respect to ${\xi}.$  In this way,  $G$ generates time coordinate transformations for solutions of Hamiltonian Einstein equations. Iff there is no time-like Killing vector field, then the solution essentially depends on time and hence exhibits  change. Thus change is exactly where it should be in Hamiltonian GR, at least for solutions of the equations of motion.  This fact might make one question the significance of definitions of observables yielding no change:  perhaps the supposed absence of change is founded on a faulty definition due to taking shortcuts?  If so, then requiring Hamiltonian-Lagrangian equivalence (henceforth $H$-$L$ equivalence for short) should solve the problem.  Thus there is a guiding theme of the series of works by Pons, Salisbury, and Shepley (\emph{e.g.}, \cite{PonsSalisburyShepley,ShepleyPonsSalisburyTurkish,PonsSalisbury}): 
\begin{quote} 
We have been guided by the principle that the Lagrangian and Hamiltonian formalisms should be equivalent \ldots in coming to the conclusion that they in fact are.  \cite[p. 17]{PonsReduce} \end{quote} 

%
%

%%%%%%%%%%%%%%%%%%%%%%%%%%%%%%%%%%%%%%%

\subsection{Should Observables Be the Same for $H$ and $L$?}

	Suppose that someone asks whether light is bent  due to gravity in GR?  A reasonable answer would be ``yes, and that was observed in 1919.''  Another possible answer is to answer with this question:  \begin{quote} Are you using Hamiltonian or a Lagrangian formalism? \end{quote}
But such a response appears to be a category mistake:  someone who understands what the word ``observables'' means in anything resembling an ordinary sense of the term (as opposed to a technical stipulation) would never imagine that the answer could depend on the choice of a Hamiltonian \emph{vs.} a Lagrangian formalism.  
Observables in any sense connected with the ordinary meaning of the term as involving  observations, should be the same in both formalisms, or more briefly, $H$-$L$ equivalent.
On the other hand, the second answer might have some justification given the role that ``observables'' play in Hamiltonian but not Lagrangian discussions. Perhaps an ``observables'' concept is instead merely a technical notion related to quantum mechanics and having little connection to observations \cite[p. 105]{Kiefer3rd}; this is a fair summary of what the typical definition of observables actually accomplishes.  Whether it reflects Bergmann's intentions is a more subtle question.  Below we will find Bergmann occupying more than one position, often in quick succession, but endorsing $H$-$L$ equivalence on a number of occasions even while systematically violating it.  
  Initially $H$-$L$ mathematical equivalence was required in the work of Rosenfeld and of Bergmann and collaborators \cite{RosenfeldQG,BergmannNonlinear,BergmannBrunings,AndersonBergmann}. After all, in reinventing the Hamiltonian formalism without a full Legendre transformation, what other standard of correctness than $H$-$L$ equivalence could there be? One needs Hamiltonian self-confidence in order for there to be independent Hamiltonian ideas. Thus such ideas would naturally have arisen only later.

%%%%%%%%%%%%%%%%%%%%%%%%%%%%%

\subsection{What First-Class Constraints Really Do in GR} 

Somewhat as Leibniz suggested (``Calculemus!''), in  subjects marred by longstanding debate, it is useful to test traditional lore with calculations whenever possible.  One intriguing piece of lore is the claim that the Hamiltonian constraint   $\mathcal{H}_0(x)$   generates both  time evolution and refoliation (change of time coordinate).
Such a claim, if true, would certainly make it difficult to find real change, because the real change supposedly generated would be only a shuffling of descriptive fluff.  Fortunately it is possible to check \emph{via} calculation.  Using some traditional results \cite[p. 241]{Sundermeyer}, one has $  \{ g_{ij}(x), \int d^3y \epsilon^{\perp}(y) \mathcal{H}_0(y) \} =    \delta^{\mu}_i \delta^{\nu}_j \pounds_{(\epsilon^{\perp} n^{\alpha}) } g_{\mu\nu}(x) $ (plus some terms vanishing with the field equations).  That looks  good, but as was noted above,  $\{ N(x), \int d^3y \epsilon^{\perp}(y) \mathcal{H}_0(y) \}=  0$ and $\{ \beta^i(x), \int d^3y \epsilon^{\perp}(y) \mathcal{H}_0(y) \}=0.$  These latter results are consistent neither with time evolution (which should involve something like a time derivative for the lapse and shift) nor with a coordinate change (which should involve something like a Lie derivative along a time-like vector field).  Hence $ \mathcal{H}_0$ generates \emph{neither} time evolution nor coordinate change, a pleasingly egalitarian result.  Rather, the Hamiltonian constraint is the  \emph{star player} on both teams. 
   Likewise for the momentum constraint $\mathcal{H}_i$ and space, one has $ \{ g_{ij}(x),  \int d^3y \epsilon^k(y) \mathcal{H}_k(y)  \} = \pounds_{\xi} g_{ij}(x) $ and  $ \{  \pi^{ij}, \int d^3y \epsilon^k(y) \mathcal{H}_k(y) \} = \pounds_{\xi} \pi^{ij}$, which look like part of a coordinate transformation or a translation, but then  $ \{  \beta^i(x), \int d^3y \epsilon^j(y) \mathcal{H}_j(y) \} = 0 $ and $ \{ N(x),  \int d^3y \epsilon^j(y) \mathcal{H}_j(y) \} = 0 $, which show that $\mathcal{H}_i$ generates neither a spatial translation nor a spatial coordinate transformation.  Rather, $\mathcal{H}_i$ is the star player on both teams, generating key terms for the spatial metric and its canonical momentum but not addressing the quantities for which the Legendre transformation failed. 
 The role of the Hamiltonian constraint and momentum constraint is thus closely analogous to the role of American athlete Bo Jackson, a star player in two professional team sports:
  %{2-sport Star Team Members:  Bo Jackson,  $\mathcal{H}_0$ and $\mathcal{H}_i$ } 
%\vspace{-.1in}
% \begin{figure}
%%%\resizebox{\textwidth}{!} 
%  { {\includegraphics[width=2.35cm,height=3cm]{BoJackson2sport.jpg}} } % or RoadConstructionFotosearch_k7631444uncomp.tiff
%%\vspace{-.1in}
%{\caption{{Bo Jackson} (pictured), $\mathcal{H}_0 $ (not pictured), $\mathcal{H}_i$ (not pictured). \hspace{1in}}}
%%\cite{BoJackson2sport}    }   } %\label{y} 
%\end{figure}
%\vspace{-.2in}
%
 Bo was an  All-Star in baseball and a Pro Bowler in American football \cite{BoJacksonWikipedia}. Such two-sport excellence is  perhaps unparalleled in the world of sports.
 Yet even Bo was not \emph{himself} a football team ($11$ players) or a baseball team ($9$ players).  Likewise $\mathcal{H}_0$ does not generate time evolution (that's team $H$) or coordinate transformations (that's team $G$).  A similar conclusion holds about $\mathcal{H}_i$ and space.  While $H$ and $G$ bear a spooky resemblance, the fact that $H$ takes $3$-dimensional space and builds space-time, whereas $G$ takes space-time and simply relabels it, makes the distinction between them clear.

%%%%%%%%%%%%%%%%%%%%%%%%%%%%%%

\subsection{Testing Definitions using Massive Electromagnetism}

Whatever one means by ``observables,'' they are the same for empirically equivalent theories (or theory formulations).  That seems to me to be a Moorean fact (in the sense of G. E. Moore's common-sense realism), more secure than anything that might be raised against it, on pain of misusing the word ``observables.''  Empirical equivalence can therefore be used to test and reject candidate definitions of observables.  A concept that is not appropriately related to ``observables'' might still be interesting or important in some other way, perhaps by pertaining to the true degrees of freedom or being useful  or essential for quantizing gravity; hence research programs attached to flawed definitions of observables might have a merely verbal rather than substantive deficiency.  

Another way to motivate the gauge generator, then, is to require equivalent observables from equivalent theories.   
    Massive electromagnetism comes in non-gauge (Proca) and gauge (Stueckelberg) versions.  The  Proca photon mass term is $-\frac{1}{2}m^2 A^{\mu} A_{\mu}.$ The mass term makes all constraints second-class:  there are nonzero Poisson brackets among the two constraints at each point.  There being no gauge freedom, everything is observable, including $A_{\mu}.$  The 
    Stueckelberg photon mass term is $-\frac{1}{2} m^2 (A^{\mu}-\partial^{\mu} \epsilon)(A_{\mu}-\partial_{\mu} \epsilon).$ The gauge compensation ``Stueckelberg field'' $\epsilon$ makes all constraints first-class, restoring gauge freedom much as in Maxwell's electromagnetism with `massless photons' (following the particle physics custom for using quantum words for physics that might or might not be quantum).   Requiring equivalent observables for these equivalent theories implies that $A_{\mu}-\partial_{\mu} \epsilon$ is observable, because it is equivalent to $A_{\mu}$ in the Proca formulation.  That result follows from using $G$ but conflicts with using  separate constraints \cite{ObservablesEquivalentCQG}.  Thus for electromagnetism the Pons-Salisbury-Sundermeyer definition of observables is vindicated \cite{PonsSalisburySundermeyerFolklore}.

%%%%%%%%%%%%%%%%%%%%%

\subsection{Should the Poisson Bracket of Observables Be 0? } 

%Having completed the survey of Bergmann's views on observables, we can profitably consider in more detail the question of whether the Poisson bracket of observables with the gauge generator for space-time coordinate transformations ought to be $0$ (part of Bergmann's usual view over the years) or nonzero, such as the Lie derivative. 

This  subsection recalls why a vanishing Poisson bracket, even with $G$, is not appropriate for observables in relation to coordinate transformations.  This impropriety follows from looking under the hood of the classical Lie derivative \cite{GRChangeNoKilling}.
  As Bergmann  himself expressed on more than one occasion   \cite{BergmannNonlinear,BergmannLectures,GoldbergConservation,BergmannHandbuch}, variations due to small coordinate transformations come in 2 varieties.  The most physically natural variation, $\delta_C$, compares field values at the same world point in different coordinate systems, like the tensor transformation law, but gives non-tensors typically (except in the case of scalars and pseudoscalars, where it gives $0$).  The most mathematically natural variation, $\bar{\delta}$, compares \emph{different} world points with \emph{same} coordinate values in \emph{different} coordinate systems.  This variation is justified not by physical meaning (which is bizarre), but by desirable mathematical properties, namely, commuting with partial differentiation  \cite{BergmannLectures,GoldbergConservation}.  The two are related by the   
\begin{quote} identity 	$ \bar{\delta}K \equiv \delta_C K - K,_{\rho}\xi^{\rho}. $''  \cite{BergmannHandbuch} \end{quote}
The difference is just the transport term.  (Sign conventions differ in this context.) 

With this distinction in mind, one can ask which variation is implemented by the gauge generator $G$.  The answer is that the $G$ generates the Lie derivative \cite{CastellaniGaugeGenerator} and hence implements the mathematically motivated variation  $\bar{\delta}$, which compares different space-time points with the same coordinate values in different coordinate systems.  The Lie derivative includes the transport term 
 $K,_{\rho}\xi^{\rho}$. % comparing \emph{two world points}: same label  in different charts.  
%   Tensor transformation law compares \emph{same}  world point in different coordinate systems, physical flavor  $\delta_C$.
Thus for the space-time metric one gets the 
%  Better chance of measuring $\#$ 5 than the Ricci scalar (!?)
  Lie derivative $ \pounds_{\xi} g_{\mu\nu} = \xi^{\alpha} g_{\mu\nu},_{\alpha} + g_{\mu\alpha} \xi^{\alpha},_{\nu} + g_{\alpha\nu} \xi^{\alpha},_{\mu}$.  The Lie derivative has terms with different origins.  From the tensor transformation rule one gets the  coordinate corrections $g_{\mu\alpha} \xi^{\alpha},_{\nu} + g_{\alpha\nu} \xi^{\alpha},_{\mu} $.
The remaining term, the transport term, here is $\xi^{\alpha} g_{\mu\nu},_{\alpha}$.
%   Quest for few gauge-invariant components $\rightarrow$ no corrections.
 Note that requiring a  $0$  Poisson bracket  makes even scalars $\psi$  constant: $\pounds_{\xi} \psi = \xi^{\mu} \psi,_{\mu} \stackrel{!}= 0$. 
  Even scalars  aren't ``observables''  in that sense \cite{TorreObservable}.  But this is not an interesting physical result showing that change is missing from the world as described by GR.  One has simply demanded the infinitesimal analog of sameness at 1 a.m. daylight saving/summer time and 1 a.m. standard time  (an hour later).  Clearly being the same at two moments an hour apart (or at two moments infinitesimally separate) has nothing to do with being observable in the ordinary sense or with being real.  The diagram shows two moments with the same coordinate values in different coordinate systems, but the toy car is nonetheless quite observable.

 \begin{figure}
%\resizebox{\textwidth}{!} 
 \includegraphics[width=5cm,height=3cm]{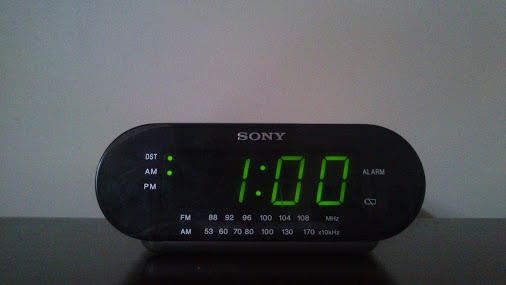} \includegraphics[width=5cm,height=3cm]{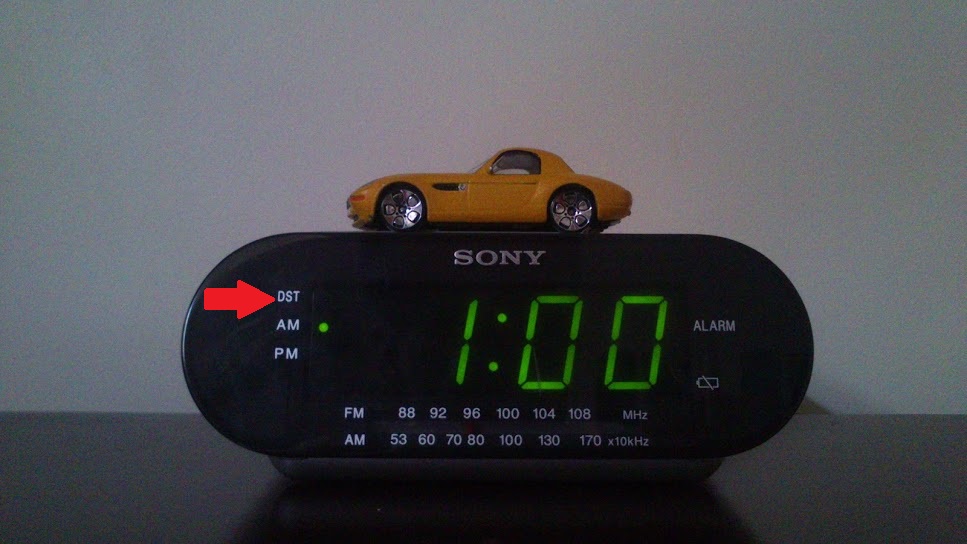} % or RoadConstructionFotosearch_k7631444uncomp.tiff
%%% \visible{ {\includegraphics[width=4.5cm,height=3cm]{PrimeMeridianCambourneCambridge2.jpg}} } % or RoadConstructionFotosearch_k7631444uncomp.tiff
\caption{{Coordinate Variations: Daylight Savings Time (left)  \emph{vs.} Standard Time} }     %\label{y} 
\end{figure} %$\bar{\delta}$-
%\vspace{-.1in}

%%%%%%%%%%%%%%%%%%%%%%%%%%%%%%%%%%%%%%%%%%%%%

%
% 
%  
%

The possibility of pointing to space-time coordinate values and coordinate transformations contrasts with internal symmetries such as electromagnetic or Yang-Mills gauge transformations and local Lorentz ($O(1,3)$) transformations of an orthonormal basis.  Because there is no way to point to internal symmetry gauges, anything observable must be invariant.  But for space-time coordinates an algorithmic conversion procedure (tensor calculus for space-time coordinates, akin to natural language translation between English and Spanish, \emph{e.g.}), suffices.  There is no obvious need for invariants, akin to propositions in Frege's heaven. It might be nice if such things exist, but plausibly GR offers no such entities, or at best offers them only in terms of \emph{physically} individuated points. But one can point to coordinate values or transformations thereof.   One can point to the Prime Meridian.  A bit west of Cambridge between the villages of Comberton and Toft one can find a sign picking out the Prime Meridian; across the street one finds the Cambridge Meridian Golf Club. Hence one can point to coordinate values with a golf course.   But one never finds oneself crossing a surface of vanishing electrostatic potential $A_0=0$; still less can one build a golf course there.  Thus one can see why invariance is necessary for internal transformations, but covariance is both possible and plausibly sufficient for coordinate transformations.

%%%%%%%%%%%%%%%%%%%%%%%%%%%%%%%%%%%%%%%%%%%%%%%%%%%%%%%%%%%%%%%%%%%%%%%%%%%%%%%%%%%%%%%%%%%%%%%%%%%%%%%%%%%%%%%%%%%%%%%%%%%%%%%%%%%%%%%%%%%%%%%%%%%%%%%%%%%%%%%%%%%%%%%%%%%%%%%%%%%%%%%%%%%%%%%%%%%%%%%%

\subsection{Testing Definitions of Observables using Massive Gravity}

Bergmann seems never to make use of the idea of  requiring equivalent observables from equivalent theories.  His arguments are all more tenuous than this principle and so, in the event of conflict, should give way to this principle.  On occasion comparison to parametrizing classical mechanics is made, a comparison that comes close but still falls short of adequate detail.  Parametrization of massive gravity leads to the conclusion that observables  should be only covariant, not invariant, under coordinate transformations, once one requires equivalent theories (or formulations thereof) to have equivalent observables \cite{ObservablesEquivalentCQG,ObservablesLSEFoP}.

Earlier it was noted that requiring equivalent observables for equivalent formulations of massive electromagnetism vindicates the gauge generator $G$ rather than separate first-class constraints.  Presumably the same result holds generally.  But a derivation involving electromagnetism might not be relevant if one has a theory, such as GR, with some relevant disanalogies to electromagnetism.  
    As massive electromagnetism comes in gauge and non-gauge versions, so does massive gravity  \cite{ObservablesEquivalentCQG}.  The graviton mass term is something like $\gamma_{\mu\nu}^2$ where $\gamma_{\mu\nu} \sim g_{\mu\nu} - \eta_{\mu\nu}.$   As the `photon mass' term breaks gauge freedom, so does a `graviton mass' term. 
One can (re)install gauge freedom with gauge compensation fields, in this case clock fields.    Observables in non-gauge massive gravity are obvious because  everything is observable: there is no gauge freedom, so no first-class constraints.

    We can therefore \emph{test} definitions of observables using two equivalent formulations of massive gravity, one without gauge freedom (everything is observable because no first-class constraints), one with gauge freedom and  first-class constraints \cite{ObservablesEquivalentCQG,ObservablesLSEFoP}.   In non-gauge massive gravity \cite{OP,FMS}, the quantity $g^{\mu\nu}$ is an observable; the coordinates are naturally Cartesian.  In the gauge version, the coordinates are arbitrary but the clock fields provide a transition to Cartesian coordinates:  $X^A,_{\mu} g^{\mu\nu} X^B,_{\nu}$.  This entity in the gauge formulation is equivalent to  $g^{\mu\nu}$  in the non-gauge formulation.  Demanding that observables in gauge massive gravity  be equivalent to non-gauge massive gravity observables implies that $X^A,_{\mu} g^{\mu\nu} X^B,_{\nu}$ is observable.  One can then simply calculate and see what $G$ does to $X^A,_{\mu} g^{\mu\nu} X^B,_{\nu}$ in order to learn how observables behave in gauge massive gravity.  The answer is that they change by a Lie derivative (making use of Hamilton's equations in some cases as needed).  But parametrized massive gravity has the same Euler-Lagrange equations has GR with $4$ scalar fields (one with a negative sign).  So surely GR observables behave in the same fashion.  Hence GR observables are entities that change by a Lie derivative under $G$: the space-time metric tensor, the connection, the Riemann and Ricci curvature tensors, \emph{etc}.
     So there is no problem of missing change in observables.     There was only a doubly flawed definition using separate first-class constraints  instead of the gauge generator $G$ \cite{PonsSalisburySundermeyer} and  assuming  $0$ Poisson bracket even for external symmetries  \cite{ObservablesEquivalentCQG,ObservablesLSEFoP}.  
  
 %%%%%%%%%
%
%
% 
%{Testing Definitions of Observables with Massive Gravities} 

$$\begin{CD}
\text{FMS Massive Gravity}  @.    \text{Parametrized Massive Gravity} \\
 \mathcal{L}: g^{\alpha\beta} \text{ observable}    @>install>gauge>  \mathcal{L}: g^{\mu\nu} X^A,_{\mu} X^B,_{\nu}  \text{ observable} \\
@.   @.  \\ 
@VconstrainedVLegendreV                                       @VconstrainedVLegendreV \\
@.      @.  \\
\mathcal{H}:  g^{\alpha\beta},  \text{ momenta }  @.   \mathcal{H}: \cancel{ \{O,FC\}=0 } \\ 
 \text{observable because }     @>demand>equivalence>                \text{or } \{O,G\}=0 \\
\text{no FC constraints }    @.                                   \text{or } \{O, G \} \sim \pounds_{\xi} O?
\end{CD}$$

 The observability of local space-time tensor fields is in some respects the fulfillment of Bergmann's intentions, which will be met below.  ``General relativity was conceived as a local theory, with locally well defined physical characteristics.  We shall call such quantities \emph{observables}.'' \cite[p. 250]{BergmannHandbuch}.  
  It turns out that observables are just $4$-dimensional   tensor calculus all over again (with Legendre transformations and  Hamilton's equations). That is also in  line with Komar-Bergmann claims that the metric is observable in intrinsic coordinates and basically any functions of the Weyl scalars can be intrinsic coordinates.

The role of this  normative introduction is justify the explicit and tacit normative claims and implicit selection criteria used in encountering and evaluating approximately everything that Bergmann (sometimes with coauthors) wrote about observables.  The goal is  to help to ascertain not merely what Bergmann said when, but which parts were justified to which degrees and what the key issues are, especially regarding separate first-class constraints \emph{vs.} the Rosenfeld-Anderson-Bergmann tuned sum $G$.   With some progress made on that front, let us now return to Bergmann's work.

%%%%%%%%%%%%%%%%%%%%%

\section{Bergmann on Observables Across the Decades} 

This lengthy section aspires to notice nearly everything that Bergmann, sometimes with students or collaborators, wrote about observables, with occasional normative remarks justified in light of the considerations above. The normative claims above also have helped to highlight which questions were asked of Bergmann's work; quite possibly there is  something interesting
 that Bergmann said about observables that is not emphasized here due to its (real or apparent) irrelevance to the normative claims.  This is, in short, internalist history of science---a tradition that seems to persist very much to the present day among historians of mathematics, I note, for whom the distinction between right and wrong results, or correct and incorrect proofs, remains important. Given that this theoretical physics is largely mathematical, it is not surprising that such normative standards might seem appropriate.  

 The period spans the early-mid 1950s to the early 1990s.  We will find a fairly stable collection of ideas, but not a coherent  view that admits reasonable examples.  Old ideas are rarely rejected, but they are often enough  supplemented by ideas incompatible with them.  Most or all of the ingredients in Bergmann's view are at least \emph{prima facie} plausible, but some have subtle defects, as the previous section on the need for the gauge generator and the  nonzero Poisson bracket for observables in GR  aimed to show.  

%%%%%%%%%%%%%%%%%%%%%

\subsection{Bergmann and Schiller (1952-53) Make  New Postulates}

In stories that start well but turn bad, it can be useful to find the first place where matters go astray.  Clearly any Hamiltonian idea that is just the Legendre transform of a sound Lagrangian idea should be a sound enough Hamiltonian idea.  On the other hand, free-standing Hamiltonian postulates have a good chance of being physically unjustified.  Hence if we find our authors admitting that they are introducing free-standing Hamiltonian ideas, we should investigate such places especially carefully.  While it is not easy (for me) to give a coherent interpretation to the whole of section 4 of the Bergmann-Schiller paper \cite{BergmannSchiller}, it does appear to be the start of novel Hamiltonian postulates. One encounters plausible motivations including  quantizing gravity and making an analogy to electromagnetism.  The latter seems both highly appropriate and not to be taken for granted from a general relativist who was motivated partly by the way that the essential nonlinearity of GR made such analogies suspect \cite{BergmannNonlinear}.  Having  argued for the value of particle physics and its use of such analogies for space-time philosophy on many occasions (\emph{e.g.}, \cite{TenerifeProgressGravity,ObservablesEquivalentCQG,LambdaMPIWG}), I especially might expect this analogy to pay off well for Bergmann.  On the other hand, another tendency among particle physicists is to deprecate analogies, picture-thinking and Gestalt `insights' (especially revolutionary ones) in favor of calculations \cite{Feynman}.  Precisely this issue comes to the fore in deciding whether one should seek gauge \emph{invariance} ($0$ Poisson bracket with whatever generates gauge transformations) or only gauge \emph{covariance} (a nonzero Poisson bracket giving a Lie derivative and hence an infinitesimal coordinate transformation) \cite{GRChangeNoKilling,ObservablesEquivalentCQG}.

For Bergmann and Schiller, there is not much pausing to worry about whether $0$ is the right Poisson bracket for a gauge transformation:
\begin{quote} 
   physical situations must be characterized by wave functionals (Hilbert
vectors) that are eigenfunctions of every constraint, belonging to the eigenvalue $0$.   \cite{BergmannSchiller}  \end{quote} 
This statement already contains both key elements of the  typical definition of observables, of having ${\bf 0}$ Poisson 
with {\bf each} first-class constraint, gently modified for a quantum context.  
Perhaps this question represents a momentary worry:
\begin{quote} 
Will the elimination of a large number of operations
from consideration not embarrass us as physicists,
by eliminating the mathematical description of
physically meaningful quantities? \cite{BergmannSchiller}
\end{quote} 
This worry is resolved without the display of any mathematics, though there is certainly a suggestion that some was done:  
\begin{quote}
In answer, we find that the observables not ruled out
are invariants, quantities that remain unchanged under
all the infinitesimal transformations with respect to
which the theory is assumed to be invariant. For
electrodynamic quantities, for instance, commutability
with the subsidiary conditions of quantum electrodynamics
implies gauge invariance.  \cite{BergmannSchiller} \end{quote}
Here again invariance ($0$ Poisson bracket) is assumed.  It is also difficult to ascertain what counts as assumption, definition or conclusion in this passage.  One might have thought that ``the infinitesimal transformations with respect to
which the theory is assumed to be invariant'' would be those generated by the gauge generator that change $A_{\mu}$ by a $4$-dimensional gradient, but perhaps then-standard practices in QED rather than Bergmann's own work on constrained Hamiltonian dynamics \cite{AndersonBergmann} are setting the standard. 
 One might also note that (Poisson) commuting with separate constraints is logically stronger than commuting with a tuned sum of them, so that such commutation implies gauge invariance but also something stronger, perhaps an insensitivity even to certain physical changes generated by separate first-class constraints.
We are assured that some mathematics has been done:   
 \begin{quote} We have gone through a number of examples to convince
ourselves that any physically observable quantity
is an invariant, but this point is so obviously of major
importance that it should be more fully investigated.
Suffice it here to say that the point of view we have
adopted is a generalization of and consistent with
accepted practices in quantum electrodynamics and
elsewhere. \cite{BergmannSchiller} \end{quote} 
If perchance  these examples were all electromagnetic in nature, then  lessons learned about $0$ \emph{vs.} nonzero Poisson brackets will not necessarily carry over if the internal \emph{vs.} external gauge symmetry distinction winds up being important---as it in fact does \cite{GRChangeNoKilling,ObservablesEquivalentCQG}, as recalled above.   
It isn't entirely obvious what either a physically observable quantity or an invariant should mean, so this process of testing examples is difficult to reconstruct.  While it is certainly true that the magnetic field and the phase space analog of the electric field are unchanged by first-class constraints, the relation between the phase space analog of the electric field (a canonical momentum) with the actual electric field (involving a velocity) involves the Hamiltonian field equations $\dot{q}= \frac{\delta H}{\delta p}$ and is imperiled by separate first-class constraints \cite{FirstClassNotGaugeEM}.  
It seems that we aren't provided the details in order to examine the grounds for Bergmann and Schiller's confidence; we can either believe them or not, or perhaps calculate ourselves and decide.  Considering how many general relativists were students of Bergmann, it is no wonder sociologically if this claim often has been believed.  Looking at Schiller's thesis, which is dated two months earlier than the Bergmann \& Schiller paper, seems not to help \cite{SchillerThesis}. %\footnote{I thank Don Salisbury for assistance in seeing Schiller's thesis. }  
  In short, motivated by quantization and analogies to QED, Bergmann \& Schiller introduced novel Hamiltonian postulates giving up mathematical $H$-$L$ equivalence, requiring constraints to act separately, and requiring a $0$ Poisson bracket (invariance not covariance).  While the $0$ Poisson bracket is doubtless correct for Maxwell, its correctness is not assured for General Relativity.  The supposed absence of change in Hamiltonian GR flows from the $0$ Poisson bracket especially, which implies a vanishing Lie (directional) derivative with respect to any vector field.  A quantity that does not change in any direction is a quantity that does not change.  Thus the origin of the supposed absence of change can be found in Bergmann and Schiller, submitted in August 1952.  If this story has a Fall after the Creation event, here it is.

%%%%%%%%%%%%%%%%%%%%%%%%%%%%%%%%%%%%

\subsection{Bergmann and I. Goldberg (1954-55) on Observables}

The paper by Bergmann and I. Goldberg work displays a continued tendency to take first-class constraints to generate gauge transformations  separately and to regard $0$ as the appropriate Poisson bracket.   
\begin{quote} 
It would be much more convincing if we could formulate
an operator algebra such that the constraints
are not only zero in the whole Hilbert space (the latter
consisting, presumably, only of permissible states) but
as a result commute with every true observable. In this
section we shall construct a transformation group in
classical phase space that has this property.   \cite{BergmannGoldberg} \end{quote} 
This paper also seems to be the first place where the problem of time is clearly manifest in print.   
\begin{quote} There are indications that the Hamiltonian of the general theory
of relativity may vanish and that all the observables
are constants of the motion.   \cite{BergmannGoldberg} \end{quote} 
It did not take long for the new premises in Bergmann and Schiller to bear fruit (good or bad).  
This is the concluding sentence of the paper and hence as such does not receive further analysis or argument.  The physics and now philosophy communities have had over 65 years to fill in the gap.

%%%%%%%%%%%%%%%%%%%%%%%%%%%%%

\subsection{True Observables in Electromagnetism (1955-56)} 

Bergmann's paper ``Introduction of `True Observables'
into the Quantum Field Equations,'' published in \emph{Il Nuovo Cimento}  \cite{BergmannObservableNC}, makes especially clear how electromagnetism provided the template for Bergmann's definition of observables. He is interested in \begin{quote} \ldots the equations of quantum electrodynamics,
with the unrestricted gauge group, that is without specializing to Lorentz or
Coulomb gauge. Even before quantization, we formulate the theory in terms
of  true observables only.    % next sentence was omitted from talks in Berlin. 
 {We define these as dynamical variables generating
canonical transformations leading from one permissible state to
another; a permissible state, in turn, is a set of values of all canonical
variables that obeys the gauge constraints.}\ldots 
 Transition to the true observables not only eliminates the longitudinal
parts, of the vector potential and the scalar potential from the
theory, but also the longitudinal components of the electric field strength. \cite{BergmannObservableNC}. \end{quote}
  Hence observables are the true degrees of freedom, cutting out all the arbitrary choices and leaving only pure physical content.  This feature gives them a form of determinism. 
 \begin{quote} The true observables are the physically meaningful variables of a theory.
Their values (at a given time) are independent of the choice of the frame of
reference (including the gauge frame). Their values can be predicted from
one time to another by integration of the canonical equations of motion (or
canonical field equations, as the case may be). Any physical situation can
be characterized uniquely in terms of the true observables. \cite{BergmannObservableNC} \end{quote} 
Thus observables are neither novel nor obscure entities:  they are the transverse true degrees of freedom, which also solve the constraints.  Who wouldn't want to find and use the general relativistic analogs, if there are any? 

 Bergmann takes these quantities to have a simple relationship to distinctively Hamiltonian concepts:  
\begin{quote}  \ldots because all the constraints \ldots  are first-class
constraints, the true observables are those dynamical variables that are left
over after we have eliminated not only the constraints themselves, but also
their canonical conjugates. More precisely, the true observables must be combinations
of dynamical variables whose Poisson brackets with all constraints
vanish. This requirement is equivalent to the one that the observables must
be gauge invariant, because the constraints are actually the generators of
infinitesimal gauge transformations.  %{textcolor{green}{ 
\end{quote} 
This last clause is a clear endorsement of the claim that any first-class constraint \emph{by itself} generates a gauge transformation.  It is indeed true that $\vec{B}[A]$ and $\vec{P}$, the phase space stand-in for $-\vec{E},$ have $0$ Poisson bracket with each first-class constraint.  It does not follow that each first-class constraint generates a gauge transformation, however.  $\vec{P}$ and $-\vec{E}$ can pick out the same observable  only if the equation relating them is preserved, and that requires using the team $G$ of first-class constraints rather than each by itself \cite{FirstClassNotGaugeEM}.  While $\vec{P}$ has $0$ bracket with each first-class constraint, $-\vec{E}$ has $0$ bracket only the team $G$:  the contributions from the primary constraint and the secondary constraint cancel due to the related coefficients (and making use of the Poisson bracket for a velocity \cite{AndersonBergmann}).  So the constraints \emph{individually} are not the generators of infinitesimal gauge transformations, even if they act that way on these phase space quantities.  (It is worthwhile to recall that those canonical momenta that do not appear in the primary constraints, can be viewed as auxiliary fields in the canonical Lagrangian $p \dot{q}-H$:  the momenta conjugate to $A_i$ are just an instance  of a standard trick in Lagrangian field theory to introduce new quantities such that their new Euler-Lagrange equations can be solved algebraically for those very quantities.  For more on auxiliary fields, see Pons \cite{PonsSubstituting}. Hence in adopting a Hamiltonian formalism one is still using a Lagrangian formalism.  The status of such momenta as mere auxiliary fields reminds us that phase space is not fundamental.)  Granting that for electromagnetism  ``true observables must be combinations of dynamical variables whose Poisson brackets with all constraints vanish'', there is no \emph{equivalence} because $\vec{P}$ has $0$ Poisson bracket with each first-class constraint, but only its transverse part is a true observable.   Thus the supposedly more precise claim is actually less precise.

Next Bergmann swerves temporarily (in the passage with the italics added) from the separate first-class constraints view back to the original team view with the gauge generator $G$.  
\begin{quote} 
\emph{ Under an infinitesimal gauge transformation
the dynamical variables transform as follows: \vspace{-.1in}  
 $$  (2) \hspace{.25in} \delta A_i =\xi,_i, \hspace{.5in} \delta \varphi = -\frac{1}{c} \dot{\xi}, \ldots \vspace{-.1in} $$  
where $\xi$  is a completely arbitrary function of the space and time coordinates
and may even depend on the dynamical variables themselves.  This transformation
$(2)$ is generated by the functional 
$$ (3) \hspace{.25in}  \mathcal{G} = - \int \bigg\{ \frac{1}{c} \pi^4 \dot{\xi} + \bigg[ \pi^s,_s + [source] \bigg] \xi \bigg\} d^3x . $$   }

It follows, then, that $\pi^4$, $\varphi,$  and the longitudinal components of $A_s$ and $\pi^s$ 
must be eliminated from the formulation of the theory.    \end{quote} 
This here-italicized passage appears to do no real work in the paper.  Its presence therefore serves to obscure the big-picture fact that he has largely  changed views from $G$ \cite{AndersonBergmann} to separate first-class constraints \cite[\S 4]{BergmannSchiller}, albeit without giving his earlier view.  
While Bergmann's views have many interesting pieces, it is hard to see how they form a coherent whole that is generally applicable.  This paper was used as a template for GR in Bergmann's best-known work on observables \cite{Bergmann}, so understanding this paper is crucial for understanding the 1961 work.  

%%%%%%%%%%%%%%%%%%%%%%%

\subsection{Observables at the Bern Conference (GR0), 1955 }

Bergmann's contribution to the 1955 general relativity conference in Bern (GR0) \cite{BergmannHPA1956} is very useful in understanding his views about observables.  This paper gets little attention, perhaps partly due to being written in German. I include a translation of parts after quoting the original passages.  Much as with the \emph{Il Nuovo Cimento} paper just discussed \cite{BergmannObservableNC} (which was the published form of a talk at Pisa that Bergmann here cites), the analogy to   electromagnetism is clearly important:  
\begin{quote} 
\selectlanguage{german} % \glqq
 Die durch die DIRACmethode ausgeschlossenen Ver\"{a}nderlichen sind solche, deren POISSONklammern mit den Bedingungen erster Klasse nicht verschwinden. Da aber die Bedingungen erster Klasse die (HAMILTONschen) Erzeugenden der invarianten Transformationen sind, so folgt, da{\ss} die noch zul\"{a}ssigen Erzeugenden Invarianten sein m\"{u}ssen.\ldots %Um Ihnen ein Gef\"{u}hl daf\"{u}r zu geben, was dies involviert, m\"{o}chte ich schnell die Ver\"{a}nderlichen \ldots %bezeichnen,  die in der Theorie des elektromagnetischen Felds durch die Eichkovarianz ausgeschlossen sind [24].  
Da die Zwangsbedingungen auf das Verschwinden der zum skalaren Potential konjugierten Impulsdichte und auf die Bestimmung des longitudinalen elektrischen Feldes durch die Ladungsdichte hinauslaufen, so folgt, da{\ss} die einzigen Erzeugenden die transversalen Anteile des Vektorpotentials (also das magnetische Feld) und des elektrischen Feldes\ldots. % \grqq  %'' % (und nur von diesen abh\"{a}ngige Funktionale) sind. Ferner d\"{u}rfen nur ganz bestimmte Kombinationen der Elektronenwellenfunktionen mit dem Vektorpotential als Erzeugende eingef\"{u}hrt werden, n\"{a}mlich solche, die eichinvariant sind.  
\cite{BergmannHPA1956}  \\
 \selectlanguage{english} %From my friend ..., slightly touched-up:
\\ The variables excluded by the Dirac method are those whose Poisson brackets with the first-class constraints do not vanish. But since the first-class constraints are the (Hamiltonian) generators of the invariant transformations, it follows that the still-permissible generators must be invariants. \ldots % In order to give you a feeling for what this involves, I would like to quickly refer to the variables that are excluded by the gauge covariance in the theory of the electromagnetic field [24].    
Since the constraints result in the vanishing of the momentum density conjugate to the scalar potential and in the determination of the longitudinal electric field by the charge density, it follows that the only generators are the transversal components of the vector potential (i.e. the magnetic field) and the electric field \ldots.\footnote{Thanks to Alex Blum for checking all translations in this section.} % (and only functionalists dependent on these). Furthermore, only very specific combinations of the electron wave functions with the vector potential may be introduced as generators, namely those that are calibration-invariant.
  \end{quote} 
One sees here a surprising claim that first-class constraints not only are gauge transformations, but even generate invariant transformations, transformations that change the Lagrangian by at most a boundary term---which is not true.

Bergmann then considers General Relativity. 
\begin{quote} Um nun wieder auf die Theorien zur\"{u}ckzukommen, die krummlinigen Koordinatentransformationen gegen\"{u}ber invariant sind, so m\"{u}ssen wir hier die Erzeugenden auf solche Gr\"{o}{\ss}en beschr\"{a}nken, die derartigen Transformationen gegen\"{u}ber invariant sind. Dies ist indes leichter gesagt als getan. Bisher ist n\"{a}mlich in der allgemeinen Relativit\"{a}tstheorie nicht eine einzige nicht-triviale Invariante bekannt. Es gen\"{u}gt ja nicht, skalare Felder zu finden; als Funktionen ihrer Argumente (der Koordinaten) transformieren sich Skalare auch. Wahre Invarianten, glaube ich, werden sich als \"{a}u{\ss}erst komplizierte Funktionale der gegenw\"{a}rtig bekannten Feldgr\"{o}{\ss}en entpuppen.    \cite{BergmannHPA1956}  \\ % \begin{quote} 
\\ Returning to the theories which are invariant with respect to curvilinear coordinate transformations, we must limit the generators to such quantities which are invariant under such transformations. This is more easily said than done. So far, not a single non-trivial invariant is known in general relativity. It is not enough to find scalar fields; scalars also transform as functions of their arguments (the coordinates). True invariants, I believe, will turn out to be extremely complicated functionals of the currently known field quantities.  %(Translated largely by my friend G\ldots) 
    \end{quote}  
 One notices that Bergmann's thinking is motivated primarily by analogy to  electromagnetism and/or by quantization, not by GR.  When he alludes to the transport term $\xi^{\alpha} \phi,_{\alpha}$ that keeps even scalars from qualifying as observables by his definition, he makes a technically correct inference.  But given that this term  arises due to comparing \emph{different} space-time points  \cite{BergmannLectures}, one wonders why requesting such a term to vanish is not seen as a  \emph{reductio ad absurdum}, as feeding in by hand an unreasonable requirement of constancy.  Physical meaning is not given its due, so the opportunity to distinguish between internal and external transformations, at least regarding  whether the Poisson bracket should vanish, is overlooked.

Bergmann is somewhat sensitive to the implausibility, if not of the definition of observables, then at least of its extension (the collection of things satisfying the definition). 
\begin{quote} 
Man k\"{o}nnte nun eine Theorie ablehnen, die in Bezug auf zul\"{a}ssige Ver\"{a}nderliche derartig {\glqq}exklusiv{\grqq }  ist. Eine solche Ablehnung erscheint mir aber voreilig. Am Beispiel der elektromagnetischen Theorie sehen wir, da{\ss} die verbotenen Gr\"{o}{\ss}en, also das skalare Potential, der longitudinale Teil des Vektorpotentials und derjenige der elektrischen Feldst\"{a}rke, entweder durch Eichtransformationen beliebiger Werte f\"{a}hig sind oder aber durch andere Zustandsgr\"{o}{\ss}en (die Ladungsdichte) bereits festgelegt sind. Diese Gr\"{o}{\ss}en k\"{o}nnen also entweder \"{u}berhaupt nicht auf Grund von Anfangsbedingungen zu einer Zeit f\"{u}r eine andere Zeit vorausgesagt werden, oder sie sind nicht unabh\"{a}ngig. Zwei formal vorgegebene physikalische Situationen lassen sich entweder als wesentlich verschieden oder aber als zwei verschiedene Beschreibungen desselben objektiven Zustands nur auf Grund ihrer Invarianten identifizieren. Ich glaube also, da{\ss} nur die im DIRAC-schen Formalismus zugelassenen Erzeugenden physikalisch als {\glqq}wahre Observabeln{\grqq} anzusprechen sind. Deshalb mu{\ss} man dieses Programm der Quantisierung sowohl formal als auch physikalisch als vern\"{u}nftig ansehen.  \cite{BergmannHPA1956} \\ \\
One could now reject a theory that is so ``exclusive'' in terms of permissible variables. Such a rejection seems premature to me. Using the example of electromagnetic theory we see that the forbidden quantities, \emph{i.e.} the scalar potential, the longitudinal part of the vector potential and that of the electric field strength, are either capable of gauge transformations to arbitrary values or are already fixed by other state variables (the charge density). Thus, either these quantities cannot be predicted on the basis of initial conditions at one time for another time, or they are not independent. Two formally given physical situations can be identified either as essentially different or as two different descriptions of the same objective state only on the basis of their invariants. So I believe that only the generators allowed in Dirac's formalism are physically to be regarded as ``true observables.'' Therefore, one must consider this program of quantization both formally and physically reasonable.  \end{quote}
Clearly much hangs on the validity of the analogy between electromagnetism and GR.  While that analogy is often useful \cite{Gupta} and might be held as a presumption, that presumption can be defeated by calculations and reflections showing that a supposed demand for gauge invariance is really a demand for constancy in the case of GR \cite{GRChangeNoKilling,ObservablesEquivalentCQG}. 
One also might wonder whether invariants are really needed for equivalence problem; is not equivalence secured if there \emph{exists} a coordinate transformation such that  $g_{\mu\nu}^{\prime} = g_{\alpha\beta} \frac{\partial x^{\alpha}}{\partial x^{\mu^{\prime}}}   \frac{\partial x^{\beta}}{\partial x^{\nu^{\prime}}}  $?

A proposal by Newman and Bergmann provided a way to work towards finding observables given the definition.  
\begin{quote} NEWMAN hat nun vorgeschlagen, diesen Invarianten durch ein N\"{a}herungsverfahren auf die Spur zu kommen, welches von der linearisierten Theorie ausgeht [27]. Wenn man die Gravitationspotentiale nach einem \emph{ad hoc} Parameter entwickelt, wobei die nullte N\"{a}herung der flache MINKOWSKIsche Raum ist, so kann man auch die Koordinatentransformationen in \"{a}hnlicher Weise in Potenzreihen entwickeln, derart, da{\ss} die auf beliebiger Stufe abgebrochene Theorie gegen\"{u}ber einer Transformationsgruppe invariant ist, die aus der Gruppe krummliniger Koordinatentransformationen dadurch hervorgeht, da{\ss} man auch deren Entwicklungen an derselben Stelle abbricht. %Von der ersten N\"{a}herung an darf man in der Transformationsgruppe willk\"{u}rliche Funktionen einf\"{u}hren. Die nullte N\"{a}herung besteht aber nicht aus der Identit\"{a}t, sondern ist die LORENTZ-gruppe.  Die erste N\"{a}herung f\"{u}r sich allein ist kommutativ und der Eichgruppe sehr \"{a}hnlich, aber nicht kommutativ zusammen mit der nullten N\"{a}herung. Die erste N\"{a}herung liefert dann genau die PAULI-FIERZschen Gleichungen f\"{u}r Gravitonen. Dar\"{u}ber hinaus ist noch nichts bekannt. Bis zur ersten N\"{a}herung l\"{a}{\ss}t sich das Programm, Invarianten zu finden und die Theorie nur mit ihrer Hilfe zu formulieren, m\"{u}helos durchf\"{u}hren. Aber erst danach wird es wirklich interessant. 
Immerhin  ist es vielleicht bemerkenswert, da{\ss}, wenn man auf FOURIERzerlegung verzichtet, die wahren Observabeln der ersten N\"{a}herung die doppelt transversalen Potentiale und ihre kanonisch konjugierten, also nichtlokale Gr\"{o}{\ss}en sind, die man durch Integrale ausdr\"{u}cken mu{\ss}.    \\  
\\ 
NEWMAN has now suggested that these invariants be tracked down by an approximation method based on linearized theory [27]. If one expands the gravitational potentials according to an \emph{ad hoc} parameter, where the zeroth approximation is the flat MINKOWSKI space, one can also expand the coordinate transformations in power series in a similar way, such that the theory which is broken off at any arbitrary point is invariant with respect to a transformation group from the group of curvilinear transformations of coordinates by breaking off their expansions in the same place.\ldots % From the first approximation one may introduce arbitrary functions in the transformation group. The zeroth approximation, however, does not consist of identity, but is the LORENTZ group. The first approximation by itself is commutative and very similar to the gauge group, but not commutative with the zeroth approximation. The first approximation yields exactly the PAULI-FIERZ equations for gravitons. In addition, nothing is known yet. As far as the first approximation is concerned, the program of finding invariants and formulating the theory with their help is effortless. But only after that does it become really interesting. 
After all, it is perhaps remarkable that, if we dispense with FOURIER decomposition, the true observables of the first approximation are the doubly transversal potentials and their canonical conjugates, that is, non-local quantities, which must be expressed by integrals.  \cite{BergmannHPA1956}  \end{quote}

He clearly recognizes the now-traditional problem of finding observables.   Constraints are used separately, not as the team   $G$. It is also clear that invariance, not covariance, is assumed to be the right behavior for observables, in that they are expected to have $0$ Poisson bracket under gauge transformations.  Such an assumption is plausible if one takes electromagnetism as the model \cite{BergmannObservableNC}, but turns out to conflict with the requirement that equivalent theory formulations have equivalent observables as applied to massive gravity  \cite{ObservablesEquivalentCQG}.

  At some point there would be work on solving the constraints of general relativity and finding true degrees of freedom \cite{DeserDecomposition,YorkDecompose}. Such work would seem to be very closely analogous to Bergmann's strand of thought involving true degrees of freedom and the transverse parts of $\vec{A}$ and $\vec{E}.$  Such work seems not to have been taken up as important for Bergmann's idea of observables in GR, but it does seem relevant to the transverse quantities defined \emph{via} integrals found in a linear approximation.

%%%%%%%%%%%%%%%%%%%%%%%%%%%%%%%%%%%%%%%%%%%%%%%%%%%%%%%%%%%%%%%%%%%%%%%%%%%

\subsection{Newman and Bergmann }

Next one might consider the work of Newman and Bergmann.\footnote{The paper by Bergmann, I. Goldberg, Janis and Newman \cite{BergmannGoldbergJanisNewman}   talks a good deal about invariant transformations (those changing the Lagrangian density by at most a divergence and hence preserving the Euler-Lagrange equations). This paper is worth mentioning here, but seems less relevant than the others and would require detailed study to identify where specific ideas about observables are presupposed. } 
One of the qualities that Bergmann expected of observables is that they be predictable in terms of the evolution equations and hence untainted by arbitrary functions of time (and place)  \cite{BergmannNewman}.    Metric components are not predictable as functions of coordinate values, because coordinate values don't even pick out space-time points without help from the metric components.  Thus one wants observables, which are expected to be predictable.  

I note that this is asking too much of GR and makes sense if one's expectations are formed by electromagnetism.  One already expects in GR, on the grounds of the coordinate transformation apparatus common in GR since the 1910s and also the point-coincidence argument \cite{HowardPointCoincidence}, that space-time points are not identifiable by mere coordinate values apart from the metric.  One also knows that many respectable quantities, such as the metric tensor $g_{\mu\nu}$, are not available in themselves, but only relative to a coordinate system with a transformation law.\footnote{An alternative, the metric-in-itself ${\bf g} = g_{\mu\nu} {\bf d}x^{\mu} \otimes {\bf d}x^{\nu},$ is not numerical and hence is remote from the language of 1950s physics. } Hence it is unclear why one would expect such quantities as these predictable ``observables'' to exist and be important.

A series expansion of the Lagrangian, equations of motion, and observables  starting around Minkowski space-time is developed.  
These fascinating and rigorous results, however, presuppose rather than test a definition of  observables.  
    It would be of considerable interest to see what this series expansion yields when observables are expected to be covariant (to have a suitable $4$-dimensional Lie derivative as their Poisson bracket with $G$) instead of the definition (requiring  invariance) used in this paper.  

% 

%%%%%%%%%%%%%%%%%%%%%%

\subsection{Bergmann at Chapel Hill 1957 (GR1), Unpublished} 

  This fairly compact but  rich third-person report is supplemented by a published product by Bergmann based on the same conference talk, so the treatment can be somewhat brief.
The notion of ``true observables'' comes up repeatedly and is held to arise whether one uses a Hamiltonian (Dirac), Lagrangian (Schwinger), or path integral (Feynman) method to quantize.  
  \begin{quote} A ``true observable'' will be described by an operator which, when applied to the state vector, produces another vector which satisfies the same constraints as the original vector.  A reduced Hilbert space is envisaged in which the only canonical transformations which are physically meaningful are those generated by true observables. \ldots BERGMANN believes that it is immaterial whether the Lagrangian or Hamiltonian approach is used to discover the non-trivial true observables; the results will be the same in either case. \cite[pp. 74, 75]{DeWitt57ChapelHill} \end{quote} 
Thus at Chapel Hill Bergmann was clearly a proponent of $H$-$L$ equivalence.  

Somewhat later Bergmann's view of the relevance of properly sorting out the classical theory as an aid to quantization came up.  
\begin{quote} WHEELER remarked that all of these discussions lead to the conclusion that the problems we face are problems of the classical theory.
\\
BERGMANN agreed, and expressed his conviction that once the classical problems are solved, quantization would be a ``walk.'' \cite[p. 92]{DeWitt57ChapelHill} \end{quote}
It is not entirely clear how such views fit with Bergmann's introducing novel Hamiltonian postulates at the classical level as an aid to quantization \cite[\S4]{BergmannSchiller}.  
But this present  project of seeking classical clarity by not introducing novel Hamiltonian postulates is somewhat encouraged by Bergmann's Chapel Hill remarks.

%%%%%%%%%%%%%%%%%%%%%%%%%%%%%%%%%%%%%%%%%%%%%%%%%%%%%%%%%%%%%
 
\subsection{Bergmann at Chapel Hill 1957 (GR1), Published}

Bergmann provided a 3-page published summary of the 1957 Chapel Hill conference \cite{BergmannChapelHill}.  While this summary was appropriately reflective of the topics discussed, it also naturally reflects Bergmann's own special interests in canonical quantization and observables. Regarding the Cauchy problem,  Bergmann noted with regret the absence thus far  of
\begin{quote} a method for constructing the variables whose values at time $t$ are determined
uniquely by the initial conditions set at the
time $t_0$. If we had such a construction, these variables
would presumably be identical with the ``true observables,''
about which I shall have more to say later. \end{quote} 
 While noting a difference of opinion regarding the need to attend to observables for path integral quantization, 
\begin{quote} 
[t]here appears little doubt that for all approaches belonging
to the [canonical or Lagrangian]  class the construction of so-called true observables is an unavoidable preliminary.

For those who are not familiar with the terminology
let me repeat that a true observable is a physical variable
whose value is independent of the choice of coordinate
system, gauge frame, and the like. Incidentally,
true observables are also the appropriate variables for
the formulation of the Cauchy problem. Analytically
they may be described as variables whose Poisson
brackets with all the constraints of the theory vanish.  \cite{BergmannChapelHill}   \end{quote} 
Again both the $0$ Poisson bracket (invariance as opposed to covariance) and the use of separate constraints is evident, as is the model of electromagnetism in providing gauge invariant transverse true degrees of freedom as a model for Bergmann's concept(s) of observables in GR.  
He then proceeded to mention  briefly the work of G\'{e}h\'{e}niau and independently  Komar on coordinates built out of curvature scalars and the work of  Newman and Bergmann 
on series expansions for observables.   

%%%%%%%%%%%%%%%%%%%%%%%%%%

 \subsection{Bergmann's 1957 Brandeis Lectures} 
 
Bergmann gave a set of not-so-well known lectures at Brandeis in 1957 \cite{BergmannLectures}, yielding a 44-page set of lecture notes by Nicholas Wheeler. The lectures include an insightful distinction between weak (trivial) and strong covariance and a clear explanation of the classical origins of the transport term in the Lie derivative formula.  Regarding the weak \emph{vs.} strong covariance distinction, Bergmann envisages the former as a formal feature of admitting the use of any coordinate system---which might involve hiding preferred coordinates, such as Cartesian coordinates in Poincar\'{e}-invariant theories, which coordinates are still preferred due to ``the rigidity of the classical metric'', and the operational significance of coordinates is known in advance---whereas the latter is a substantive feature of lacking preferred coordinate systems and having \emph{a priori} unknown operational significance of the coordinates, as GR presumably exemplifies (see also (\cite{StachelGC}).  These lecture notes also give a very clear discussion of the transport term in the Lie derivative as arising by comparing \emph{different} space-time points with the \emph{same} coordinate values in \emph{different} coordinate systems.  Hence arises $\bar{\delta}$ variation, which is motivated mathematically by commutation with partial derivatives.  Bergmann and his school had discussed these matters clearly before  and would do so again  \cite{BergmannNonlinear,GoldbergConservation,BergmannHandbuch}.  While it seems to me that this clear understanding of the origins of the transport term should have been a \emph{reductio ad absurdum} of the idea of observables as \emph{invariant} under coordinate transformations for exactly the same reasons that an object can be observable even if its properties differ between 1:00 a.m. standard time and 1:00 a.m. daylight saving/summer time, as a historian one can at least note that ignorance about the transport term played no role for these authors in facilitating the conception of observables as invariant rather than covariant.  

% RosenfeldQG on transport term,  Weyl 

The treatment of observables follows from an insightful consideration of the identity of world points, the Cauchy problem, efforts by Komar to use curvature scalars as coordinates, and the question whether two solutions are equivalent.  The way that observables are introduced shows that Bergmann clearly intended for observables to be related to observations.  He says that mathematics becomes physics ``only when one is able (i) to point to specific quantities
and expressions in the formalism and designate them as `observable' and (ii) to
prescribe operational procedures by which such quantities may, in fact, be
measured (observed).'' Hence any notion of observables that is utterly remote from observations---\emph{e.g.}, one that implies that observables do not change or require integration over the entire universe---is at odds with this intention of Bergmann's.  He sought a definition that was not empty and yet did not depend on a particular physical theory; a theory itself indicates what its observables are.  Taking the view that only predictable quantities are worth trying to predict,  ``we separate from the class of all dynamical variables (associated with a given theory) the subclass of those which are predictable and designate them as `observable.'{''} (p. 27)
This move is not necessarily harmless:  if in fact nothing, or nothing humanly accessible, is predictable in the sense of a Cauchy problem, then Bergmann's move discards all (humanly accessible) relevant physical quantities and retains  chimeras (or perhaps unicorns \cite{KucharCanonical93}).  If one can set up a Cauchy problem only by gauge-fixing, then one gets as many Cauchy problems as there are gauge (coordinate) fixings; gauge fixings that agree this week could still diverge next week \cite{Bergmann}.  What if nothing is predictable \emph{simpliciter}, but everything is predictable up to coordinate transformations?  Why is that not good enough?

Bergmann was fully aware of the shortage of known observables in GR.  
\begin{quote} In Einstein's general relativity \ldots not a single non-trivial observable is
known at present.\ldots Interestingly even among the
trivial observables not one is local. It seems plausible that all observables
in general relativity are non-local. \cite[p. 27]{BergmannLectures} \end{quote} 
This shortage did not, however, inspire doubt about the definition.
The fact that the Hamiltonian formalism has thus far not been mentioned implies  $H$-$L$ equivalence.

The introduction of the Hamiltonian question of observables is, not surprisingly, in terms of Maxwell's electromagnetism.  
Indeed Bergmann feels able to provide a general definition of observables on this basis. 
\begin{quote} 
 For our electrodynamic experience and intuition make it clear
that observability in Maxwell's theory is equivalent to gauge-invariance. Now the gauge-invariance of a quantity $F$ can be expressed as the condition that the
Poisson brackets with the constraints, $[F, \Pi^0]$ and $[ F, {\bf \nabla \cdot \Pi}] $,\ldots vanish. [Actually the vanishing of these brackets is logically stronger than gauge invariance, which requires only that the two terms cancel when combined with coefficients of which one minus the time derivative of the other \cite{FirstClassNotGaugeEM}.] 
This last formulation, however, clearly allows of more general application. We
shall, in fact, henceforth adopt it as our criterion of observability. The
observability problem then becomes the problem of discovering a complete set
of (Poisson) commutators with the constraints.
\cite[p. 29]{BergmannLectures}  \end{quote} 
But  vanishing is the right thing for the Poisson brackets to do only to the degree that all theories' gauge freedom is relevantly similar to that of Maxwell's electromagnetism.  For internal gauge symmetries that seems fair enough.  For the coordinate transformations of GR, however, Bergmann's earlier treatment of the Lie derivative and its transport term, alas, show that this analogy fails, though Bergmann did not notice.  
Thus one sees both the presence of the separate first-class constraints doctrine and the vanishing Poisson bracket doctrine in this passage.  Thus these doctrines, though superficially plausible, have a  slender logical foundation.

What becomes of $H$-$L$ equivalence?  While we have seen that Bergmann believed in it during these lectures, he has introduced a second definition of observables (p. 29), a distinctively Hamiltonian one, inequivalent to the one provided on p. 27. (Equivalence would hold if the separate constraint and vanishing Poisson bracket doctrines were true.) 
Bergmann has by no means given up belief in $H$-$L$ equivalence, however.  
  ``For those dynamical systems to which both formalisms [Hamiltonian and Lagrangian] are relevant
they are, of course, equivalent. This suggests that the observability criteria already outlined in the Hamiltonian context have also a natural Lagrangian formulation.'' \cite[p. 33]{BergmannLectures}
Reliance on the Poisson bracket might seem to show otherwise.
But Bergmann takes this problem to have been overcome by Peierls and his own school \cite{BergmannSchiller,BergmannGoldbergJanisNewman}.

%%%%%%%%%%%%%%%%%%%%%%%%%%%%%%%%%

\subsection{Bergmann and Janis on Subsidiary Conditions} 

This 1958 paper explored the use of coordinate conditions (gauge fixing) in covariant theories such as GR, akin to Fermi's treatment of electromagnetism.  The result seems to be that coordinate conditions do not hurt---the same observables emerge from different coordinate conditions---but coordinate conditions do not help fundamentally, either. \begin{quote} We define observables as functions (or functionals)
of field variables that are invariant with respect to coordinate
transformations. Physically, they are the only  quantities that lend themselves to observations in
(conceptual) experiments that are constructed within
the conceptual framework of the theory.  \cite{BergmannJanisSubsidiary} \end{quote}
Invariance under coordinate transformations is curiously weak as a definition of observables. 
Given Bergmann's notion of invariance,  $A_{\mu} g^{\mu\nu} A_{\nu}$, a scalar (albeit changing under electromagnetic gauge transformations), is not a problematic case:  it isn't invariant because of the transport term.  But what is to prevent a quantity from being invariant under coordinate transformations but variable with the electromagnetic gauge, such as $\int d^4x \sqrt{-g}  A_{\mu} g^{\mu\nu} A_{\nu} $ integrated over all space-time?   This might be a slip rather than a reliable indicator of their Bergmann and Janis's views, however.

A more substantive treatment comes later.  ``We call true observables of a classical (i.e., non-quantum) 
theory those quantities that are in principle measurable within the theory.'' Clearly the intention here is to tie observables to observations. 
\begin{quote} 
 In order to make the
concept of measurability more precise, let us say that
the value of a true observable at a time $t$ can be predicted
(at least in principle) from a sufficient set of
data at an earlier time $t_0.$ [Quantum footnote suppressed] Otherwise a quantity whose
behavior is completely unpredictable within the framework
of the theory is nevertheless measurable; but
then one would search for a more complete theory, in
which the quantity in question would also be predictable.  \cite{BergmannJanisSubsidiary} \end{quote} 
In my view it is unclear why anything needs to be predictable full-stop, rather than just predictable up to coordinate transformations.  It is quite true that any statements about parts of the future that are not infinitesimally nearby---perhaps the value of some stock market in one year's time, or perhaps something of more cosmological interest---involve implicit reference to the metric tensor in order to define proper times, as Bergmann has carefully explained \cite{BergmannLectures}.  That is somewhat complicated, but it seems to be just how things are in GR and is manageable; indeed numerical relativists successfully manage it routinely. 
Consider a linguistic analogy:  there are contexts in which one cannot predict the language in which multi-lingual speakers will continue the conversation.  That would be true even if one knew somehow the proposition (perhaps regarding the whiteness of snow) about to be expressed. Perhaps ``Snow is white'' will be uttered, or ``Schnee ist wei\ss,'' or some other equivalent.  
 But sufficiently competent listeners are able to `translate' from the sentence to the proposition or a  sentence in the listener's preferred language or the language of thought or some such and thus get by, despite the lack of predictability of the word-sounds.  (This analogy accounts for the non-transport terms in the Lie derivative, the terms arising directly from the tensor transformation rule.) 
 If nothing is predictable in the stronger sense,  then there are no true observables in the sense stipulated.  Should one seek a more complete theory?  It seems to me that GR lacks predictability in the strong sense mentioned and yet gets by just fine due to determinism \emph{modulo} coordinate transformations.  Seeking a more complete theory is required only if one is confident that an adequate theory should have a property that GR apparently lacks.  So we cannot rule out the possibility that ``true observables'' in this sense do not exist, especially if we aim to understood GR just as it is, without \emph{ad hoc} postulates adding in further desiderata.  Given that Bergmann was attracted to canonical quantization partly in order to avoid suppressing distinctive features of GR, there is irony here.

Employing as a premise that there is a predictable true observable $A$, Bergmann and Janis  infer that one can ``construct a constant of the motion whose value is equal to
the value of $A$ at the time $t_0$; if we restrict ourselves to consideration of only those quantities which are constants
of the motion, we lose no information about the true observables.'' This statement might be true and, if true, is important.  But one cannot be confident that it is true from the argument given.  The reasons are similar to those besetting Aristotelian Scholasticism and Wolffian metaphysics:  one encounters superficially impressive and apparently rigorous long chains of reasoning leading to substantive knowledge claims, but often in retrospect something is amiss (perhaps  with the premises or with terms that fail to refer) and the conclusions are not very  reliable.  Hence the progress of knowledge in constrained Hamiltonian dynamics requires separating lore from theorems following from secure premises.

%%%%%%%%%%%%%%%%%%%%%%%%%

\subsection{Bergmann \& Komar at Royaumont 1959 (GR2) }

	The paper in the proceedings from the Royaumont GR conference displays both many familiar themes and a stronger emphasis on intrinsic coordinates than most of Bergmann's previous work, which is natural given Komar's work on intrinsic coordinates.  
Bergmann by now was quite confident of a mathematical definition of observables based on Poisson brackets: 
``\emph{observables}, quantities which in the Hamiltonian formalism I shall define as dynamical 
variables (field variables or functionals of the field variables) whose unmodified  Poisson brackets with all the constraints of
the theory vanish.'' \cite{BergmannKomarRoyaumont} 
One sees again both the use of  separate first-class constraints (as opposed to the gauge generator $G$) and a $0$ Poisson bracket invoked for all gauge freedoms, without attention to the internal \emph{vs.} external distinction.  Despite Bergmann's intentions, this definition is essentially Hamiltonian, whereas a definition in terms of the gauge generator  which would be compatible with $H$-$L$  equivalence.  
This definition of observables is explained in terms of a supposed feature of the first-class constraints.
\begin{quote} 
 In the Hamiltonian formalism the  constraints are the generating densities of infinitesimal coordinate transformations. 
 Our definition of observables, that they be dynamical variables whose Poisson brackets with all constraints vanish, amounts to the requirement that the observable be invariant with respect to (infinitesimal) coordinate transformations. \cite{BergmannKomarRoyaumont} \end{quote} 
Note again the use of separate constraints and the claim that they generate coordinate transformations.
For Bergmann and Komar, being invariant is not exemplified by a scalar field.
\begin{quote} 
In order to avoid misunderstandings, permit me to explain my terminology.  A \emph{scalar} is said to be a quantity whose value at a fixed world point remains unchanged under coordinate transformations.  An \emph{invariant}, on the contrary, is a quantity so defined that its value in every coordinate system is the same; that is to say, it may be an invariant integral, or it may even be a function of the coordinates or of some other parameters, but defined so that whenever we calculate its value, it comes out the same regardless of the coordinate system in which we perform the calculation.  \cite{BergmannKomarRoyaumont} \end{quote} 
    While Bergmann's work emphasized Hamiltonian methods, he (and Komar) continued to affirm $H$-$L$ equivalence (whether or not the definition of observables offered satisfied it). 
``At any rate it is certainly worthwhile to explore the possibilities inherent in the Lagrangian formalism.  Our definition of observables as invariants is already independent of the formalism.\ldots'' 

%   
% Elaboration added in response to referee's request.  

Fittingly in light of Komar's coauthorship, there is much calculation involving ``intrinsic coordinates,'' functions of  Weyl scalars: $4$  scalar fields made of quadratic or cubic expressions in the Weyl curvature tensor. Such entities are useful for individuating space-time points, a task that vexed Einstein during his hole argument phase and that ultimately requires some kind of relational story. (For a recent discussion perhaps unexpectedly preferring a passive viewpoint, see (\cite{WeatherallHole}).) Weyl curvature scalars, at least when sufficiently many of them are independent (which holds in the absence of Killing vector fields or the like), exhibit in an invariant (scalar) way how features of the metric distinguish one space-time point from another. Rather than values of $g_{\mu\nu}$  or some concomitant thereof \emph{relative to some coordinate system}, leaving the further task of deciding (in)equivalence based on the (non)existence of a coordinate transformation with certain features, one can  use any convenient coordinate system to calculate the (scalar) values of these functions of the Weyl tensor.

 Bergmann and Komar  claim to achieve local observables, including the metric in intrinsic coordinates (see also \cite{KomarObservable}).  This is a striking claim in view of the previous belief that no local observables were known.  Bergmann and Komar would not maintain this claim too much longer, however, it seems.  

Something rather like Bergmann and Komar's claim emerges if one takes the view that observables should be covariant rather than invariant \cite{ObservablesEquivalentCQG,ObservablesLSEFoP}. The fact that any coordinate system is \emph{some} functions of Weyl scalars if one of them is, tends to deflate the interest in using Weyl scalars (or functions thereof) as coordinates, however.

%%%%%%%%%%%%%%%%%%%%%

\subsection{Bergmann \& Komar Local Observables from Weyl Scalars (1960)}

Here Bergmann and Komar think enough of their finding local observables using Weyl scalars to publish the result in \emph{Physical Review Letters} \cite{BergmannKomarPRL}.    Somehow the same quantities are supposed to be both scalars and coordinates. Scalars being independent in value of any choice of coordinates, and coordinates being a matter of choice, one might be wary.  Having one's cake and eating it seems to be  crucial to justifying the technical manipulations.  As befits the arbitrary nature of coordinates, arbitrary functions of the original Weyl scalars are also admitted.  One might wonder whether this gives up the game:  now all coordinates are intrinsic if any coordinates are, and one will need a coordinate transformation rule relating results from one (intrinsic) coordinate system to another.  Invariance of observables is the doctrine, but covariance seems to be emerging.  While in my view this was actually progress, in any case there is a kind of internal tension that might not have helped this view to persist.  Judging by the later obscurity of this work, its  claims were not widely accepted in a context that accepted the invariance of observables.  One also notes Bergmann's own later criticisms \cite{BergmannHamiltonJacobiMixed,BergmannFoundationsGRBunge,BergmannGeometryObservables}.  ``This approach is anything but aesthetic, or even practical, and in my opinion will serve at best only as an example guaranteeing existence.''  \cite{BergmannGeometryObservables} Thus it is reassuring to know that the Weyl scalars are there, but actually using them is another matter.  It appears that if one works very hard, one can use them after all \cite{SalisburyRennSundermeyerHamiltonJacobi}.

 %%%%%%%%%%%%%%%%%%%%%

\subsection{Observables in General Relativity (1961)}

This paper is the main source typically used to learn Bergmann's views about observables.  Bergmann's  main problem to be addressed is that  in GR, one doesn't know today what tomorrow's  coordinates will mean \cite{Bergmann}, a problem that also appeared prominently in his Brandeis lectures \cite{BergmannLectures}. More explicitly, one cannot  predict $g_{\mu\nu}(t,x^i)$ by specifying $t,x^i$ and integrating some partial differential equations.  To be a bit more explicit than Bergmann, adding  a `year' to $x^0$ and holding $x^i$ fixed  doesn't mean that one sits still and waits a physical year;  one might be at some other time or place, depending on how the space-time coordinates in the intervening period relate to physical distances---that is, depending on the metric tensor.  Unlike Poincar\'{e}-covariant theories,  in which specifying a reference frame once suffices for all time, in GR the equally natural choices are so many that  today's choice does not persist:  it doesn't answer tomorrow's question, \emph{etc.}  
  While local field values are not strictly predictable, one can retain a form of  ``causality'' (determinism) by taking an equivalence class under coordinate transformations \cite{Bergmann}. One might call this a covariant story, as opposed to an invariant one.  

 Bergmann does not rest with a covariant story, however, but seeks to find genuinely predictable invariant quantities. Footnote 7 indicates that the paper on observables in electromagnetism \cite{BergmannObservableNC} provides the background.  He then starts defining observables.  
 ``We shall call a quantity an observable if it
can be predicted uniquely from initial data.'' \cite{Bergmann}. Observables are thus not tainted by tomorrow's `free' choice of coordinates with \emph{a priori} unknown metrical properties. 
Defining observables is too important a task to do just once, however.  Bergmann immediately continues: 
\begin{quote} This definition is complete, but not always the most
useful. A somewhat modified definition would be to
call an observable a quantity that is invariant under a
coordinate transformation that leaves the initial data
unchanged. \end{quote}   This second definition seems to lose sight of electromagnetic gauge freedom, unfortunately, which is surprising given how often  $\vec{E}$ and $\vec{B}$ have been taken as paradigms. 
One notices that \emph{active} diffeomorphisms, a source of renewed confusion from modern geometry after they befuddled Einstein during his hole argument phase, are nowhere in sight, because Bergmann's understanding is still purely passive. (In my view that is a virtue, because passive transformations do not assume that one can separate points from their properties. Hence mathematical points are not dissimilar to physical points.) Note also that so far,  $H$ hasn't appeared.  ``Observables'' thus far are not a Hamiltonian concept, so $H$-$L$ equivalence holds.  
Some authors have seized upon these passages and found therein a definition of ``Bergmann observables,'' but the diversity of Bergmann's thoughts on the subject and his frequent endorsement of what others call ``Dirac observables'' (having $0$ Poisson bracket with each first-class constraint) make the term ``Bergmann observables'' deeply misleading, at least insofar as the goal is to understand Bergmann's view.  If the goal is instead to understand GR, then other concerns have been raised above.

%%\begin{figure}
%%%\resizebox{\textwidth}{!} 
%%{\includegraphics[0in,0in][2.1in,1.35in]{}} % or RoadConstructionFotosearch_k7631444uncomp.tiff
%%\caption{Fixing Coordinates:  The Job that Never Goes Away in GR}  %\label{ConeHyper} }
%%\end{figure}
%
%\begin{figure}
%%\resizebox{\textwidth}{!} 
%  { {\includegraphics[width=11.843cm,height=6.136cm]{BergmannCoordinatePicture.png}} }  \vspace{-.3in}    %  911 by 472 % or RoadConstructionFotosearch_k7631444uncomp.tiff
%%%% \visible{ {\includegraphics[width=4.5cm,height=3cm]{PrimeMeridianCambourneCambridge2.jpg}} } % or RoadConstructionFotosearch_k7631444uncomp.tiff
%{\caption{{Fixing Coordinates:  The Job that Never Goes Away in GR}    }   } %\label{y} 
%\end{figure}

At this point Bergmann introduces now-familiar Hamiltonian claims.  
\begin{quote} 
If we employ a Hamiltonian formalism to describe our theory, then there is a certain set of generators that produce infinitesimal coordinate transformations.  An observable is then a dynamical variable that has vanishing Poisson brackets with all the generators of infinitesimal coordinate transformations. This last
definition is rather useful: The generators of infinitesimal
coordinate transformations are known in closed
form. \cite{Bergmann} \end{quote}    Unfortunately there is no such ``set'' of generators, because  individual constraints do not generate gauge transformations in typical examples \cite{FirstClassNotGaugeEM}; first-class constraints rather generate gauge transformations when combined to form the gauge generator $G$  \cite{CastellaniGaugeGenerator}.  If one assumes that Bergmann's views form a coherent package, then this statement, being sufficiently mathematical that one can calculate with it, is far more interesting (``useful'') to the physicist (not the philosopher)  than the earlier verbal definitions, which the physicist will pass over quickly.  Hence one might easily not notice that this mathematical definition is inequivalent to the earlier verbal definitions.  

Another difficulty with finding an alleged ``dynamical variable that has  vanishing Poisson brackets\ldots'', I notice, is the tacit assumption that one is likely dealing with a handful of components, easy for taking Poisson bracket. That is true for  electromagnetism.  But in GR, we already know what is predictable, and it is not a handful of components.  Two things predictable from initial data are the metric-in-itself ${\bf g}=  g_{\mu\nu} {\bf d}x^{\mu} \otimes {\bf d}x^{\nu}$ (if one has modern tastes) and the geometric object $\{g_{\mu\nu}\}$, the components in \emph{all} coordinate systems \cite{Nijenhuis,Anderson} (if one has classical tastes)---both assumed to be evaluated at a \emph{physically} individuated point (a metric-dependent notion). Probably anything else that is predictable (such as a curvature scalar) is constructed out of such ingredients.  But  ${\bf g}=  g_{\mu\nu} {\bf d}x^{\mu} \otimes {\bf d}x^{\nu}$ is rather abstract and non-numerical, whereas  $\{g_{\mu\nu}\}$ in all coordinates has high cardinality; neither is an obvious candidate for taking Poisson brackets.

Bergmann explores the role of intrinsic coordinates built from quadratic and cubic contractions of the Weyl tensor.  
 \begin{quote} 
The four expressions $A^1 \cdots A^4$ exhaust
the scalars that can be formed from the metric tensor
at this differential level. Thus, they recommend themselves
by their relative simplicity, though later developments
may lead to a preference for other scalars. Any
set of four functions $f(A^1 \cdots·A^4)$ may now serve as
intrinsic coordinates.

The components of the metric tensor in the chosen
intrinsic coordinates are observables. Moreover, they
form a complete set, as knowledge of the metric is
obviously sufficient for determination of all properties
of our manifold.  \end{quote}  
As noted above, these statements go very far toward the author's view that observables should be covariant as opposed to invariant.  Given that the four given contractions of powers of the Weyl tensor do not even all have dimensions of the same power of length, the ability to accommodate transformations to other functions of contractions of Weyl scalars---one might say, a coordinate transformation rule---is evident.

Finally Bergmann explores the possibility of splitting the general relativistic coordinate freedom into a gauge-like part analogous to electromagnetism and an  asymptotic Lorentz part. He anticipates that invariance under the gauge-like part and covariance under the Lorentz part might be appropriate.  The (non)existence of such a split is closely related, he proposes, to the (non)existence of a gauge-invariant gravitational energy.     Bergmann had previously written a brief but fascinating paper on gravitational energy \cite{BergmannConservation} still worthy of attention today.

%%%%%%%%%%%%%%%%%%%%%%%%%%%%%%%%%%

\subsection{Lorentz and Gauge-Like Transformations?} 

In this paper Bergmann explores in more detail the possibility of dividing general relativistic coordinate freedom, at least in cases with suitable boundary conditions, into a gauge-like part and a Lorentz-like part \cite{BergmannLorentz}.   He seeks invariance under gauge-like transformations and only covariance under such asymptotic Lorentz transformations. 
Bergmann makes some specific technical assumptions and arrives at a negative result, while acknowledging that other technical assumptions might lead to different results.  This attempted split occupies more or less the same conceptual territory as the later  Bondi-Metzner-Sachs group \cite{BondivdBMetzner,SachsBMS}, a topic of renewed discussion in physics recently.   He provides some definitions of observables.  
    ``By definition, observables are quantities that are individually independent
of the choice of coordinate system but whose values depend on properties of a particular
solution of the field equations of general relativity.'' This somewhat unfamiliar wording is linked to the Bern paper \cite{BergmannHPA1956}.  Later on the same page he makes another statement which seems to be consistent with the earlier one.  
  ``Thus intrinsic coordinates are naturally associated with the construction of
observables, by definition quantities that will not change their values under any coordinate transformations.''  \cite{BergmannLorentz}
Both of these discussions are too brief to make clear how observables are intended to differ from scalars, matters on which Bergmann is more expansive elsewhere, as we have seen.

A natural question to ask about intrinsic coordinates is what to do with symmetric space-times, in which the number of independent contractions of the Weyl (or more generally the Riemann) tensor dwindles.  One might hope that as symmetries of the metric (Killing vectors or the like) grow and independent scalars shrink in number, the former will somehow compensate for the latter.  Bergmann reports that ``Kerr has shown that in the absence of Killing fields there always exist locally defined scalar fields that may be ·used as a complete set of intrinsic coordinates.''  That as-yet unpublished work presumably corresponds to one or both of Kerr's rather obscure papers on the topic \cite{KerrScalarsMotionsPositive,KerrScalarsMotions}; see also Komar's work \cite{KomarScalars}. % According to MathSciNet (which covers some physics literature as well as mathematics), each paper has been cited only three times, one of which is shared for a total of five citing papers.  

%%%%%%%%%%%%%%%%%%

\subsection{\emph{Handbuch der Physik} Article (1962)} 

Are observables evident in observations?  Recent evaluations of the typical formulaic definition of observables point out that observables have precious little to do with observations  \cite{SmolinPresent}  \cite[p. 105]{Kiefer3rd}.  Yet we have seen in Bergmann's Brandeis lectures show that he clearly intended for observables to be observable. That same intention is clear in his little-read 1962  encyclopedia article.   
%   Kiefer:  no.  ``It must be emphasized that there is no a priori relation of these observables to observables in an operational sense.'' \cite[p. 105]{Kiefer3rd}  
  ``General relativity was conceived as a local theory, with locally well defined physical characteristics.  We shall call such quantities \emph{observables}.''   \cite[p. 250]{BergmannHandbuch} 
His  habit of defining the term ``observables'' repeatedly is evident.  
Later on the page one gets another definition.  \begin{quote} We shall call \emph{observables} physical quantities that are free from the ephemeral aspects of choice of coordinate system and contain information relating exclusively to the physical situation itself.  Any observation that we can make by means of physical instruments results in the determination of observables; because  by assumption all coordinate systems are in principle equivalent, a physical measurement by its very nature is either empty (that is to say, we could have predicted the result of the observation without any specific knowledge of the physical situation), or it provides us with some physical information.  It cannot provide us with information about the coordinate system, because the coordinate system is a matter of free choice, for us to make without any restriction caused by the characteristics of a particular physical situation.
 \cite[p. 250]{BergmannHandbuch}  \end{quote} 
 These two definitions and the two definitions in his widely read paper \cite{Bergmann} all seem to be compatible.  All point toward local geometric objects (either components-in-all-coordinates or boldfaced objects-in-themselves). 
The metric tensor  and matter  components everywhere in space-time in some coordinates (satisfying the field equations)  would ``include all the physical characteristics of the situation to be described'' \cite[p. 250]{BergmannHandbuch}.  That claim contrasts with the view that  only mysterious non-local observables, the ``deep structure,'' are real and can sustain real change \cite{EarmanMcTaggart}. 
According to Bergmann,  $g_{\mu\nu}$ and matter fields in all coordinates would be merely  \emph{redundant} (not unphysical) in that (1) Cauchy data at some initial moment should suffice to imply the whole of space-time (obviously some topological assumptions are thereby made) and (2) some of the component information is  just coordinate (gauge) artifacts, not real physics  \cite[p. 250]{BergmannHandbuch}. But redundancy doesn't imply non-reality. 
Bergmann's commitment to observables as local and tied to observations shows clearly that one cannot consistently attribute to Bergmann any notion of  observables that includes or implies a lack of change, nonlocality or globality. But somewhat like one making a  list of desired Christmas presents, Bergmann wanted a whole lot of attractive features to characterize observables, not all of which can be had together in GR even if most or all can be had together in Maxwell's electromagnetism.

%%%%%%%%%%%%%%%%%%%%%%%%

\subsection{Bergmann-Komar, Warsaw \&  Jab{\l}onna 1962 (GR3)} %Status Report

This conference paper with Komar \cite{BergmannKomarStatus} pays considerable attention to the question of observables in the service of quantum gravity.    ``In this article we shall review primarily
one facet of this search [subjecting GR to field quantization], the present status of the search for \emph{observables}.''  \cite[emphasis in the original]{BergmannKomarStatus} 
Prior to discussing observables, they mention a weaker notion of  equivalence under coordinate transformations.  ``We shall call two \emph{solutions equivalent} if they can be carried over into each other by means of a coordinate transformation. Likewise, we shall call two permissible \emph{fields equivalent} if they lead to equivalent solutions.''   \cite[emphasis in the original]{BergmannKomarStatus}

Such traditional and consensual notions of tensor calculus are soon followed by 1950s Hamiltonian ideas.   Surprisingly given both that aim and Bergmann's tendency to define observables repeatedly in the same work, one sees little  in the way of definition in this paper. One does quickly encounter a claim that (presumably separate) first-class constraints generate gauge transformations.
\begin{quote}
Within the compass of these definitions, infinitesimal mappings
generated by linear combinations of the constraints (with variable
coefficients) map permissible fields on equivalent fields. The constraint
generators thus enable us to cover the whole constraint hypersurface with mutually exclusive \emph{equivalence classes} of permissible fields.
These equivalence classes are the points of a new phase space, the \emph{reduced phase space}. \cite[emphasis in the original]{BergmannKomarStatus} 
\end{quote}  
I note several  worries  about this construction of a reduced phase space \cite{GRChangeNoKilling}.  First, given that spatial coordinate transformations are generated by the spatial gauge generator \cite{CastellaniGaugeGenerator} and not by separate first-class constraints, it is not true that the points thus identified are all related by spatial coordinate transformations.  A similar point holds for temporal changes of coordinates, but with the further complication that one needs to make use of Hamilton's evolution equations as well  \cite{FradkinVilkoviskyHLEquivalence}.  Furthermore, temporal coordinate transformations are  
velocity-dependent and live not in phase space, but in phase space  $\times$ time  \cite{MukundaSamuelConstrainedGeometric,SuganoGaugeGenerator,SuganoGeneratorQM,LusannaVelocityHamiltonian,GRChangeNoKilling}---one might call it phase space-time.  While moving from mathematics  to casual words is relatively common and relatively harmless in physics, this is an example of moving from casual words to mathematics, a quite different matter.  
Bergmann and Komar infer that quantities defined on this reduced phase space are ``constants of the motion'' in a sense including 
\begin{quote}  both the usual requirement that the variable in question does not change
its value in the course of time, but also the further requirement that
it is invariant with respect to coordinate transformations. In view
of the fact that in general relativity the change in the course of time
is defined only up to a coordinate transformation, only invariant
functionals can be constants of the motion in every possible coordinate
system. Our result is then that constants of the motion (used
as generators) map equivalence classes on each other, that is to say,
they generate infinitesimal canonical transformations in the reduced
phase space. \cite{BergmannKomarStatus} \end{quote} 
Given the flawed premises, this argument is unsound.  One might also think that vanishing Lie derivative along some time-like vector field would be a natural candidate for defining constancy, but that criterion is not employed.

Coordinate conditions (gauge fixing) are explored, including Anderson's treatment of electromagnetism that restricts the $3$-vector potential and its conjugate momentum to be transverse.  
Intrinsic coordinates are defined in a somewhat surprising way.  ``If there is precisely one possible coordinate system left in each equivalence class, we shall
call that coordinate system `intrinsic.' '' \cite{BergmannKomarStatus}  Hence being intrinsic in this sense has more to do with completeness of the gauge fixing, and less to do with coordinates that are suggested by the space-time geometry, than usual. 
A version of the claim that intrinsic coordinates (in some sense) lead to local invariances (observables) appears:
\begin{quote} 
 If it is possible to specify a coordinate system in such a manner that in every Riemann-Einstein manifold
there is exactly one (four-dimensional) coordinate system having
the properties specified by the chosen coordinate conditions, then
every component of any geometric object at a specified world point
in that coordinate system has an invariant significance; that is to
say, its numerical value is determined uniquely for each solution,
regardless of the coordinate system in which this solution is
originally presented. \cite{BergmannKomarStatus}  \end{quote}  
A way to set up intrinsic coordinates locally involves building scalars from the Weyl tensor, at least assuming the absence of symmetries of the metric (generically true but false for all exact solutions, they note).  
Inferences are drawn about time-dependent solutions, constants of the motion, and intrinsic coordinates.  It would be interesting to explore to what degree such inferences hold when the above-noted problems with the construction of reduced phase space are addressed.

%%%%%%%%%%%%%%%%%%%%%%%%%%%%%%%%%%%%%%%%%%

\subsection{Physics and Geometry (1964/1965)}

This interesting paper was aimed at philosophers of science at the conference on the Logic, Methodology and Philosophy of Science, Jerusalem, 1964
\cite{BergmannPhysicsGeometry}.  GR implies difficulties with point individuation in terms of coordinates.  Kretschmann's critique of general covariance is discussed (though the name is consistently misspelled as ``Kretschmar''); Anderson's efforts to answer are mentioned without much specificity. (Such work was soon published   \cite{AndersonCoordinates,Anderson}.)  To address such issues it is helpful to specialize one's vocabulary, Bergmann says, so that  
\begin{quote} meaningful statements may be constructed. There exists a subset of physical
variables, the ``observables'', whose values are independent of the choice
of coordinate system employed. Thus, any relationship between observables
is ``meaningful'', and conversely, these are the only relationships that are
legitimate. A program aiming at the identification and systematic exploitation
of the observables has been under way for many years, but its execution
is hampered by profound technical difficulties, which have not yet been
overcome completely.
\cite{BergmannPhysicsGeometry}  \end{quote} 
While there are few new ideas here, it is striking how much Bergmann's subject matter suited the philosophy of science at a time when most of the serious philosophical engagement with General Relativity was long past (Reichenbach) or just beginning \cite{GrunbaumChronometry}.

%%%%%%%%%%%%%%%%%%%%%%%%%%%%%%%%%%%%%%%%%%%%%

\subsection{Radiation and Observables (1965) }

This little-known conference paper both presents some largely novel reflections on observables in general and reflects on what observables should mean in the case of asymptotically flat space-times in light of the BMS group  
\cite{BergmannRadiationObservables}. 
He recalls why he is not satisfied with tensor components or even scalars: 
\begin{quote} 
By performing arbitrary co-ordinate transformations we may give almost
any value we want to the component of a tensor, for instance, or even
to a scalar, at a world point identified by the values of its four spacetime
co-ordinates, without thereby providing any substantive information
about the physical situation in which this value occurs. 
\end{quote} 
Point individuation is key once more.

Bergmann reminds his audience of his distinction between scalars and invariants:
\begin{quote} 
the only quantities that ought to possess expectation
values are invariants, that is to say quantities that have the same
values regardless of the choice of (conventional) co-ordinate system.
Because of a frequent confusion in terminology let me say, parenthetically,
that I shall distinguish between invariants and scalars. A scalar
is a one-component field defined on the four-dimensional space-time
manifold which obeys the infinitesimal Lie-transformation law
$$ (1)  \hspace{1in}  \delta^* U = -U,_{\rho} \delta x^{\rho} $$
whereas an invariant obeys the law $\delta^* U = 0.$
\end{quote} 
One might quibble by pointing  out that not only scalars, but also pseudoscalars have this Lie derivative, because the formula for Lie differentiation depends only on coordinate transformations near the identity \cite{SzybiakLie}. The desired result $\delta^* U = 0$ is of course unlikely to accommodate local fields; Bergmann's talk of constants of motion (including the next paragraph) perhaps makes room for nonlocality, though he  at times expects observables to be local \cite[p. 250]{BergmannHandbuch}, as we have seen.  
This passage makes especially clear Bergmann's commitment to observables that are invariant and not merely covariant under gauge transformations.  This commitment will come under a bit of pressure later in the paper, however.

Because of arbitrary changes of time coordinate,
\begin{quote} 
an invariant is necessarily a constant of the motion; and thus we concluded that an observable in general
relativity is \emph{ipso facto} a constant of the motion. In a Hamiltonian formalism
all properties of the gravitational field can be expressed in terms
of appropriate Cauchy data on a space-like three-dimensional hypersurface;
Dirac has shown that the twelve field variables $g_{mn}$, $p^{mn}$ are
suitable as Cauchy data provided they satisfy at each point of the hypersurface
the four constraint conditions $H_s = 0$, $H_L = 0$. 
\end{quote} 
This claim about Cauchy data is somewhat surprising given that the Hamiltonian field equations still depend on $g^{00}$ or $g_{r0}$ or some similar set of quantities (such as the Arnowitt-Deser-Misner lapse function $N$ and shift vector $\beta^{i}$ \cite[Ch. 21]{MTW} \cite[Appendix E]{Wald}) that, with the spatial metric $g_{mn},$ permit the recovery of the space-time metric $g_{\mu\nu}.$ Without such quantities besides the spatial metric, one cannot infer the time derivatives of $g_{mn}$ even given the canonical momenta $p^{mn}.$   Without the space-time metric one cannot calculate proper distances and times, $ds^2.$   
See also (\cite{SalisburySyracuse1949to1962}).

Let us resume where we left off.  
\begin{quote} 
Obviously, the value of any constant of the motion is contained in principle in these
Cauchy data, though the data are not by themselves invariant: aside
from intrinsic data they contain information about the location of the
initial-value hypersurface and about the choice of three-dimensional
co-ordinates on that hypersurface. But we may conclude that any observable,
or constant of the motion, may be represented in principle as a
functional on a three-dimensional hypersurface. An observable may then
be characterized as a functional that has vanishing Poisson brackets with
all of Dirac's constraints; this is because the constraints are the generators
of the infinitesimal co-ordinate transformations. 
\end{quote} 
Thus we find the claim that separate constraints generate gauge transformations, at least a hint that the relevant constraints are only $\mathcal{H}_s$ and $\mathcal{H}_0$ given the lack of mention of primary constraints trivializing the momenta conjugate to $g_{00}$ and $g_{i0}$ and the earlier invocation of Dirac's phase space with only 12 field components, the relation between the definition of gauge transformations and the definition of observables, and (again) the requirement that observables should be invariant (not covariant).

Bergmann proceeds to 
 \begin{quote}
discuss the changes in the concept
of observables that suggest themselves in the restricted program that
focuses attention on a special class of solutions of general relativity that
we might call the radiative solutions. I am referring to the techniques
that were initiated by H. Bondi and his coworkers, which have been
elaborated and further clarified, among others, by E. T. Newman and
his coworkers, by R. Penrose, and by R. K. Sachs.   
\end{quote} 

``What is the appropriate
definition of an observables in the presence of an invariance group of
the type''  identified by Bondi, Metzner and Sachs, with some Lorentz-like transformations and ``supertranslations''? 
Maxwell's electromagnetism is naturally a test case from which one might draw lessons. 
``Intuitively, and guided by all our experience of many years, we should
say that an observable must be gauge-invariant, but that it need not
be Lorentz-invariant.''  
One key difference between Lorentz transformations and electromagnetic gauge transformations  is that 
``every conceivable Lorentz transformation will change the values of my Cauchy data; but there are gauge
transformations that will leave them unchanged.''
So what is the key distinction that permits mere Lorentz covariance but requires electromagnetic gauge invariance?
\begin{quote} 
From the foregoing, it appears that the chief difference between Lorentz
transformations and gauge transformations is that the latter involve
arbitrary functions of the time co-ordinate, whereas the former do not.
It is not decisive that one set of transformations is described by a finite
set of parameters (ten in the case of the inhomogeneous Lorentz group),
the other by functions.
\end{quote} 
This is surely one of the plausible candidate answers.  Another plausible candidate answer, but which Bergmann does not consider, is that electromagnetic gauge transformations are internal, whereas Lorentz transformations are external.  
His conclusion is stated tidily. 
\begin{quote}
To summarize our argument: observables must be invariant with
respect to all transformations that involve arbitrary functions of the
time but need not be invariant with respect to other transformations
belonging to the invariance group of the theory. \cite{BergmannRadiationObservables} \end{quote} 
Using considerations about factor groups and normal subgroups in application to the Bondi-Metzner-Sachs group (here described using Sachs's term ``generalized Bondi-Metzner group''), ``I believe that from the
point of view of the general theory which I have presented, any set of
variables that form a geometric object under the GBM group should
be considered observable event though noninvariant under supertranslations.''  This claim is intended not as a matter of vocabulary only, but as a guide to quantization.  He makes the inference that  Bondi's news functions will be observables, though in the discussion Komar is not convinced.

%%%%%%%%%%%%%%%%

\subsection{Hamilton-Jacobi and Sch\"{o}dinger Theory (1965) }

This paper explores a Hamilton-Jacobi formulation and Schr\"{o}dinger quantization for GR \cite{BergmannHamiltonJacobi}.
One key ingredient is Bergmann's high view of the simplification of canonical GR achieved by Dirac \cite{DiracHamGR,SalisburySyracuse1949to1962}.
\begin{quote} 
Basic to this presentation is the canonical version of
general relativity perfected by Dirac.

\hspace{.5in} {\bf II. HAMILTON-JACOBI FORMALISM} 

For what follows it is essential that in its canonical
version general relativity may be presented in terms
of a set of 6+6 canonical field variables, $g_{mn}$ and $p^{mn}$,
respectively, ($m$, $n$= 1, 2, 3) obeying Hamiltonian field
equations with a Hamiltonian that is composed of the
constraints of the theory ($\mathcal{H}_s,$ $\mathcal{H}_L$ in Dirac's notation),
with coefficients that are arbitrary functions of the
four space and time coordinates and which are algebraically
related to the remaining components of the
metric tensor, $g_{0n},$ $g_{00}.$\ldots

The constraints generate infinitesimal coordinate transformations.  Specifically, an integral of the form 

$$ C= \int d^3x[ \eta^s(X) \mathcal{H}_s + \eta^L(x)\mathcal{H}_L] \hspace{.5in}  (2.1)$$
generates an infinitesimal coordinate transformation,
in which the three coefficients $\eta^s$ represent the infinitesimal
coordinate shifts within the space-like hypersurface $x^0=$constant, whereas $\eta^L$ is the magnitude of the infinitesimal time-like displacement normal to that hypersurface.  
\end{quote} 
Unfortunately this treatment of  $g_{0n}$ and $g_{00}$ is at best ambiguous.  While  it is true that $g_{0n}$ is arbitrary partly due to the freedom to let the spatial coordinates slide around over time relative to the future-pointing normal vector built from the space-time metric and the time coordinate hypersurfaces, and it is true that $g_{00}$ is also more or less arbitrary (though it at least should remain positive, given that its `trivial' value is $1$ rather than $0$), these freedoms exist \emph{only within the context of the $4$-dimensional tensor transformation rule}  \cite{GRChangeNoKilling}.  One can choose $g_{0n}$ at will, but only while making compensating changes in the rest of the space-time metric in accord with $4$-dimensional tensor calculus.  Note also that the constraints  $\mathcal{H}_s$ and $ \mathcal{H}_L$ have $0$ Poisson brackets with the lapse function and shift vector (or whatever quantities besides the spatial metric one takes as primitive in building the space-time metric) for just  the same reason that $x$ and $p_y$ have vanishing Poisson brackets in simpler mechanical theories:  there simply are no pairs of conjugate variables involved.  (Depending on how one chooses the primitive variables, one could perhaps find some $g_{mn}$ mixed in, but the resulting contribution would still vanish on the constraint surface.)  Hence it is not the case that 
$\mathcal{H}_s$ generates spatial coordinate transformations.  Still less is it the case that  $ \mathcal{H}_L$ generates a time-like displacement normal to the hypersurface.  Whether one means a time-like coordinate transformation or time evolution,  $ \mathcal{H}_L$ is only the star player, not the team.  This would have been a good time to notice the need for the field equations (as was done later \cite{FradkinVilkoviskyHLEquivalence}) at least in the $3+1$ formulation)  and to re-reinvent the gauge generator  $G$  in $3+1$ form (as was done later \cite{CastellaniGaugeGenerator}).   When Anderson and Bergmann first reinvented the gauge generator $G$ \cite{AndersonBergmann} (after Rosenfeld's original invention \cite{RosenfeldQG,SalisburySundermeyerRosenfeldQG}), it contained velocities, a feature and/or bug that plausibly makes the result less dependent on field equations to equate phase space functions with velocities.  
It would be interesting to explore to what degree the questionable premises actually influence to results derived.

%%%%%%%%%%%%%%%%%%%%%%%%%%%%%%%

\subsection{Foundations Research in Physics (1967)}

This little-known work \cite{BergmannFoundationsPhysicsBunge} fittingly served as the first chapter in the first volume of Mario Bunge's 4-volume series \emph{Studies in the Foundations, Methodology, and Philosophy of Science}.    As has often been the case, the identification of world points, a problem no longer soluble in terms of space-time coordinates in GR as it was for earlier theories, motivates Bergmann's reflections.  
The metric is less readily measured than was taught in his younger days.  
\begin{quote} 
 In the end we realized that world points might be identified in terms of local curvature; thus the metric tensor might be replaced by the gradient fields of the curvature. We have called such involved structures observables.
\end{quote} 
This is of course the coincidence or intrinsic coordinate approach to observables, in which evidently Bergmann had not yet lost faith as of 1967, though there are warning signs.  His ongoing remarks help to explain what problems he took observables to solve and hence why he spent so much energy pursuing observables by various means despite the curious results sometimes obtained: 
\begin{quote} 
 we consider that we have related a mathematical object, the metric field, to something that can in principle be observed if we have succeeded in describing the gravitational field in terms of observables. In this description we have eliminated all references to an arbitrary coordinate system, retaining only elements that would remain unchanged if we replaced the metric field in one coordinate system by the equivalent metric field expressed in terms of a different coordinate system. Thus, in terms of observables, two formally different fields will be in fact inequivalent. Does that mean that we have succeeded in reducing the theory of gravitation to its ``guts'' elements, that we have obtained a language that tells us directly how to measure any observable by a specific experimental procedure? Far from it. We believe that we
have overcome an essential road block in our efforts to separate redundant from essential formal elements in the theory of gravitation, incidentally at a terrific sacrifice in mathematical simplicity. But though I believe that the construction of experimental procedures for the determination of the values of specified observables has been reduced to a technical question, I may eventually turn out to be wrong. We have, as it were, peeled one layer off the logical onion, and our eyes are beginning to tear. We do not know much about the layers that are left.
\end{quote}  
But one might doubt, I suggest, whether understanding the classical theory requires any notion of invariant observables; why not just use tensor calculus?  By analogy, human life proceeds well with communication in languages and (with some effort) transformations between languages.  \emph{Must} there be invariant propositions in Plato's or Frege's heaven lying behind (above? below? in, with and under?) utterances in languages?  Perhaps there must, but one should not accept such claims without carefully weighing the arguments offered  \cite{PropositionsStanford}.  In any case one sees why Bergmann thought it was important to find observables and why he discussed the matter so often.

%%%%%%%%%%%%%%%%%%%%%%%%%%%%%%%%%%

\subsection{The Riddle of Gravitation:  A Popularization}

Bergmann devoted a chapter to observables in a popular book \cite[pp. 201-205]{BergmannRiddle}.  While this chapter understandably seems to contain no new ideas, the fact that much of the verbal content can be so readily aimed at a popular audience is an indicator of the richly conceptual/philosophical nature of many of his views  on the topic as well as the importance that he placed upon it. 

%%%%%%%%%%%%%%%%%%%%%%%%%%%%%%%%

\subsection{The Sandwich Conjecture}

This little-known conference paper \cite{BergmannSandwich} does not explicitly address the issue of observables, but nonetheless is relevant.  It uses a toy model not unlike the Baierlein-Sharp-Wheeler action for GR that eliminates the lapse function as an auxiliary field from the Lagrangian (which one can do if the spatial Ricci scalar is nonzero); that paper is citation 1.  With the lapse eliminated, one will of course have no primary constraint conjugate to it, so the gauge generator (had Bergmann sought it explicitly) would have been the Hamiltonian-like constraint which he does use.  
The treatment of reduced phase space is critiqued by Salisbury, Renn and Sundermeyer \cite{SalisburyRennSundermeyerHamiltonJacobi}.

%%%%%%%%%%%%%%%%%%%%%%%%%%%%%%%%

\subsection{Status of Canonical Quantization (1969/1971)}

This paper arose from a conference held in Israel in 1969 in honor of Nathan Rosen's 60th birthday \cite{BergmannRosen}.
The paper is quite substantial and unusually carefully argued in certain respects, including the quantization procedure.  The example of parametrized nonrelativistic particle mechanics plays an important role, exemplifying the role of artificial gauge freedom and aiming for quantization that is equivalent between the original lean formulation and the plumper version that treats time as a physical variable evolving with respect to some other quantity $\theta.$

The geometry of space-\emph{time} is downplayed to a perhaps stronger degree than before.  It is difficult to find explicit mention of the four components of the space-time metric that are in addition to the spatial metric $g_{mn}.$  Still less do their canonical conjugates, which vanish as primary constraints, receive any mention.  The Hamiltonian is an integral of the sum of $H_L$ and $H_s$ multiplied by some quantities.  One does encounter certain mysterious quantities that correspond to the ADM lapse function and shift vector (though not named) if one knows what to expect, but for some time they seem to be mere loci of arbitrariness.
\begin{quote} 
The coefficient of $H_L$ is the measure for the normal distance between the initial and the
follow-up surfaces, whereas the coefficients of the three $H_s$ relate the sideways
displacements of points with identical coordinate values on the two surfaces. \cite{BergmannRosen} 
\end{quote} 
While these descriptions are recognizable as the lapse and shift \cite{MTW}, one gets little hint that such quantities are in some sense dynamical variables, that such quantities are needed in the space-time interval $ds^2$ for proper distances and times, or that their arbitrariness is intimately bound up with arbitrariness in the spatial metric $g_{mn}$ (and its canonical momenta) in accord with $4$-dimensional tensor calculus.  
\begin{quote} 
The full covariance of the theory is expressed by the lack of restrictions
on the coefficients of the four constraints in the Hamiltonian, permitting
arbitrary normal and sideways displacements in the progress from one three-surface
to the next.  \cite{BergmannRosen}
\end{quote} 
The truth in the neighborhood of this claim would seem to be that the covariance \emph{beyond manifest spatial covariance} is so expressed (to the degree that it can be one when coordinate is singled out as time)---except that one must be sure to transform the whole space-time metric together in accord with $4$-dimensional tensor calculus; here the flexibility in the lapse and the shift vector are highlighted.  One could imagine using either roughly spherical or roughly Cartesian coordinates with no distinction apparent in the lapse (a spatial scalar) or, \emph{if the shift vector vanishes}, even in the shift vector, so clearly the spatial covariance is not fully expressed in terms of the lack of restrictions on the lapse and shift.    
The notation used later for $H= \int d^3x (\xi^s H_s + \xi^L H_L)$  further downplays the lapse and shift, in that $\xi$ is a typical letter for a descriptor of an arbitrary gauge transformation, not a canonical coordinate, in Bergmann's work.  Such attitudes are clearly very far indeed from seeking to re-reinvent the Rosenfeld-Anderson-Bergmann gauge generator in $3+1 $ form.  The answer to that question (not concerning oneself with boundary terms) is, for spatial coordinate transformations of space-time, 
\begin{eqnarray}
G[\epsilon^i, \dot{\epsilon}^i] = \int d^3x [\epsilon^i \mathcal{H}_i  +  \epsilon^i \beta^j,_i p_j - \epsilon^j,_i \beta^i p_j  + \epsilon^i N,_i p + \epsilon^i,_0 p_i]
\end{eqnarray} 
if one uses $N$ and $\beta^i$ for the lapse function and shift vector, respectively; changes of time coordinate involve a somewhat similar expression involving $\mathcal{H}_0$, though one needs Hamilton's equations as well  \cite{CastellaniGaugeGenerator}.  One notices the key roles played by the momenta conjugate to the lapse and shift (the primary constraints  $p$ and $p_i$ respectively), the inclusion of the lapse and shift among the canonical variables, and their sharp distinction from the descriptors $\epsilon^i.$ 
While now-standard textbook treatments of GR in $3+1$ form give little thought to space-\emph{time} covariance \cite[chapter 21]{MTW} \cite[appendix E]{Wald} as something that a canonical formalism could have, they do not go as far as Bergmann in rendering that question inconceivable due to his high estimation of Dirac's late 1950s contribution, which eclipsed his own earlier work that aimed to preserve space-time covariance in some form \cite{AndersonBergmann}.

 Passing over the treatment of parametrized mechanics, which deserves more careful study, one finds some illuminating remarks on observables.
\begin{quote} 
First, it appears to me clear, even from a classical point of view, that the
observables of any consistently covariant theory will not be local in the conventional
sense. This is because one requires a rigid, non-dynamic metric
structure of space-time to identify a point within a frame of reference that
is determined by data lying outside the quantum dynamics of the field.\ldots  Conceivably world points will be
identifiable by means of ``coincidences''. That is to say, a world point will be
determined by means of the numerical values that certain scalar fields assume
there. Such a determination will not be unique from a global point of view,
and the same method of identification will not work for manifolds that are
not sufficiently generic, but, most important, such identification will depend
on the choice of properties.  \cite{BergmannRosen}
\end{quote} 
Point individuation is still a key problem for Bergmann, but coincidences and curvature scalars seem less clearly a solution to the problem than in times past.

%%%%%%%%%%%%%%%%

\subsection{Hamilton-Jacobi with First- and Second-Class Constraints}

What becomes of a Hamilton-Jacobi formulation if one \emph{partially} fixes the gauge with coordinate conditions?  This little-known but prize-winning work takes up that question \cite{BergmannHamiltonJacobiMixed}.  
Why might one wish to fix the gauge only partially?  
\begin{quote}  The maximal restriction of the choice of coordinate system entails the introduction of
intrinsic coordinates, which turn all constraints into second-class constraints,
thus obviating the need for a Hamilton-Jacobi treatment. [footnote suppressed] Although intrinsic
coordinates lead, in principle, to a complete set of observables in general relativity,
their defects, of which the most glaring is their gross deviation from
Lorentz coordinates, render this procedure illusory. It appears preferable to
retain coordinate systems that are approximately, or asymptotically, Lorentzian
and hence do not destroy one's intuition. 
 \cite{BergmannHamiltonJacobiMixed}. \end{quote} 
Thus fully fixing the coordinate gauge leads to coordinates simply too far removed from the familiar.  This judgment is an indication of falling out of love with intrinsic coordinates.

%%%%%%%%%%%%%

%%%%%%%%%%

\subsection{Foundations Problems in General Relativity (1971)}

This little-known work in the last volume of a series edited by Bunge \cite{BergmannFoundationsGRBunge} contains several noteworthy passages.  One is an especially clear statement of the role of the ``spirit'' of GR in guiding theory construction and choice for GR-exceptionalists.  ``If one accepts the spirit of the theory, there can be no such thing as a classical background metric, though there will be (non-local) constants of the motion.''
A bit of healthy collective self-criticism follows:
\begin{quote} 
In spite of sustained efforts, a definitive and generally accepted quantum theory of gravitation does not as yet exist. Opinions vary as to what features such a theory should possess in order to be universally acceptable. Some serious workers consider the whole program of quantization unsound.\ldots  If the gravitational field is to be quantized along with all other fields, opinions differ as to whether such a theory should possess an invariance group of the scope of the group of curvilinear transformations, or whether it should be merely Poincare-invariant. Finally, it is not at all clear whether the topological aspects of space-time are to be quantized.\ldots   Most of the workers in this field, including the present author, have strong but differing opinions on all of these points; the very fact that there is disagreement would seem to indicate that so far these opinions should be regarded as just that, not as universally agreed-to principles. \cite{BergmannFoundationsGRBunge}
\end{quote} 
The fate of the intrinsic coordinates/coincidences approach to observables is revealed.  
\begin{quote} 
Relatively little actual calculations [\emph{sic}] have been done. One calculation, which was not approximate but based on the notion of observables by intrinsic coordinates (also known as coincidences), has resulted in Poisson brackets between components of the metric tensor so involved that intuitive interpretations were not practicable. \end{quote}
 This unfortunate outcome doubtless helps to justify the later remark (already noted above) that this approach is mostly valuable as an existence proof.

% also field dependent coordinate/gauge transformations discussed.  observables, discrete spectrum and old quantum phase integral.  

%%%%%%%%%%%%%%%%%%%%%%%%%%%%%%

 \subsection{Bergmann and Komar 1972 on Transformation Groups}

Bergmann continued to regard this paper, entitled ``The Coordinate Group Symmetries of General Relativity''  \cite{BergmannKomar}, as important in later years, judging by his later citations.  This judgment seems  justified.  The paper aimed to distinguish three different gauge groups (using the term loosely---Bergmann and Komar do not trouble with the considerations that frequently attract the term ``groupoid'' as more accurate) that are often lumped together as coordinate transformations.  What are the consequences for the concept of observables?  On this point there is not much innovation:  they speak of 
``\ldots the observables of gravitation theory, that is, the functionals of the field variables that are invariant under
space-time mappings. In our view such observables are required if one is to construct a quantum theory of gravitation.'' \cite{BergmannKomar}
The presence of three distinct groups might conceivably lead to three distinct notions of observables.  Fortunately the groups have the same orbits, so observables are unified after all.  The paper includes various 
fascinating technical results about various groups (or non-groups) of transformations. 
It appears that two prominent ideas in this paper, the non-introduction of canonical conjugates of the lapse and shift and the use of active diffeomorphisms, are quite inessential to the paper's technical accomplishments.  That is good because both ingredients contribute to the conceptual problem of supposedly missing change in GR \cite{GRChangeNoKilling}.

A key contribution involves the canonical representation of changes of time coordinate:  
\begin{quote} 
it is well known that the Lie derivative of the displacement vector, representing an infinitesimal
coordinate transformation (or an infinitesimal mapping) with respect to
another displacement vector, in general involves derivatives off the initial
hypersurface, a fact that appears to foreclose the representation of the
commutator algebra in terms of functions, or functionals, restricted to a
three-dimensional domain. This apparent paradox will be resolved in this
paper. \end{quote} 
In non-canonical $4$-dimensional geometry, the commutator of two coordinate transformations (involving Lie derivatives) is just the Lie derivative with respect to the commutator.  While that is a lovely property, it does introduce the time derivatives of the descriptor vector field (in the usual sense of the term) and of the canonical variables.  While the Legendre transformation lets one eliminate velocities of the canonical variables insofar as the Legendre transformation succeeds, one achieves a Legendre transformation only for the spatial metric, not the whole space-time metric, in GR.  It appears that Bergmann and Komar have somewhat blurred the distinction between descriptors and dynamical variables (also noted in relation to (\cite{BergmannRosen}), probably because the lapse and shift vector or the equivalent are now said not to be canonical variables and so might seem more like descriptors.  But even electromagnetism employs the time derivative of the arbitrary scalar function describing gauge transformations.  The problem is the appearance of time derivatives of the lapse and shift vector or the equivalent; if one desires to express the transformation using data on just a Cauchy surface, then one is bound for frustration.  While there seems to be little or no retrospective reflection on the Anderson-Bergmann gauge generator \cite{AndersonBergmann} and why it no longer suffices, that fearsome expression clearly depends on the velocities of some canonical variables (at least canonical variables by the standards of 1951).
 Presumably the Anderson-Bergmann gauge generator seemed not adequately canonical after all.  Dirac's $3+1$ split with spatial and normal projections \cite{DiracHamGR} avoids  the time derivatives of the lapse and shift or equivalent \cite{SalisburySundermeyerEinstein,PonsSalisburyShepley,PonsSalisburyShepleyYang}, making the transformations projectable to phase space.  

   It is no surprise to find the claim that $\mathcal{H}_0$ and $\mathcal{H}_i$  generate coordinate transformations, neglecting $g^{00}$ and $g^{0n}$ or their equivalent (such as the lapse and shift) despite talking about them. Those other $4$ pieces of the space-time metric are explicitly said not to be canonical variables (p. 20), a statement that makes  clear that no canonical momenta conjugate to them are envisaged.  Hence the primary constraints are abolished, kicking away the ladder after reaching the roof.  
     
 In light of the paper's title, one might anticipate a passive viewpoint, but in fact one sees a strong leaning toward an active viewpoint of mapping points to other points, not a passive viewpoint of repainting labels while leaving world points alone.  I have not discerned any strong reason for this shift internal to Bergmann and Komar's project.  Sociologically it is perhaps worth noting that  Anderson's book had started talking (verbally) about manifold mappings, though the mathematics appears to be straight classical tensor calculus \cite{Anderson}.  Presumably this shift was in the air for those whose working language was differential geometry and who worked on a topic, General Relativity, that had long been heavily populated by mathematicians.  
Bergmann had long been interested (as shown above) in the problem that world points could not be identified in GR in terms of their coordinates; he viewed coincidences (picking out points in terms of, say, values of curvature scalars) as one viable solution.  One can hardly fail to notice that this succession of reasoning closely  parallels Einstein's hole argument  and then Einstein's point-coincidence argument as the solution to the problem posed by the hole argument.  It is thus quite unclear why Bergmann would readily adopt the notion of active transformations, dependent as they are on the idea that point individuation is primitive rather than relative to physical properties.  
With active transformations in view, Bergmann and Komar arrive at a philosophically interesting conclusion upon reflection on field-dependent transformations. 
``In addition this paper will make precise the assertion that in general relativity the identity of a world point is not preserved under the theory's
widest invariance group.''   
Perhaps one can say that \emph{if} one entertains the idea of primitive point individuation, then reflection on the widest gauge `group' undermines the idea.  If one does not entertain primitive point individuation, as one might have expected from Bergmann's earlier views, then it seems unclear how the introduction of field-dependent coordinate transformations makes a difference.  One then has both explicit and implicit (\emph{via} $g_{\mu\nu}$) dependence of the descriptor vector field (components) $\xi^{\mu}$ on space-time location, and the components of the descriptor vector field in the `same' coordinate system vary from world to world even given the same explicit spatio-temporal dependence.  But trans-world identity is at best problematic in GR anyway, making the idea of the same coordinate transformation in different worlds obscure. It seems (to me) that whatever relational views one adopted in embracing Einstein's point-coincidence argument are still appropriate.    
 
%%%%%%%%%%%%%%%%%%%%%%%%%%%

  % Peter G. Bergmann and Art Komar. Invariant noncanonical mappings of singular theories. Physical Review D, 9(12):3301{3303, 1974.
% doesn't mention observables, not obviously relevant

%%%%%%%%%%%%%%%%%%%%%%%%%%

\subsection{Geometry and Observables (1974/1977)}

This brief work shows again Bergmann's engagement with philosophers of physics  \cite{BergmannGeometryObservables}.  
    Gauge theories are formally indeterministic.
\begin{quote} 
If some of the dynamical variables are not predictable from Cauchy
data, one might conclude that such a theory is noncausal. This conclusion
appears unpalatable because (a) some dynamical variables remain predictable
(all those that are gauge-invariant), and (b) the imposition of gauge
conditions, which by assumption do not modify the physical characteristics
of the theory, render it formally causal with respect to all dynamical
variables. A way out, and the one that I have adopted, is to say that in a
theory with a gauge group no Cauchy data fix the frame of reference (the
gauge frame), but that in those theories that we are concerned with,
Cauchy data do fix the physical situation. This formulation implies that
only gauge-invariant dynamical variables are physically significant; the
inference is that only gauge-invariant quantities are susceptible to observation
and measurement by physical instruments. This conclusion is warranted
if all physical interactions, including those with physical instruments,
are necessarily gauge-invariant.  \cite{BergmannGeometryObservables}
\end{quote} 
One might wish for a stronger case that (a) is not vacuous.  One might also wonder whether observation  and measurement by physical instruments might involve inherently conventional elements and thereby yield merely covariant rather than invariant results.  

The definition of observables is not a surprise. 
    ``\ldots I reserve the term \emph{observables} for gauge-invariant dynamical variables.'' 
   His formulation continues a trend of blurring time evolution and coordinate transformations \cite{BergmannRosen,BergmannKomar}: 
``What makes the group of diffeomorphisms peculiar is that the mapping
of one  Cauchy hypersurface on another is not separable from the other
gauge transformations.''  The distinction is indeed not very clear if one discards the momenta conjugate to the lapse and shift or other equivalent of $g_{0\mu}$, fails to re-reinvent the gauge generator after adopting a $3+1$ split, and employs active diffeomorphisms---none of which is very compelling relative to Bergmann's \emph{early} work \cite{AndersonBergmann}.  Once again 
\begin{quote} 
\ldots all observables are constants of the
motion, i.e., their values are the same on all conceivable Cauchy surfaces.
This result has been dubbed ``the frozen formalism''. Its adoption is unpalatable
to many, as it appears to eliminate from the formalism all
semblance of dynamical development.

In its defense I would make two points. First, problems in ordinary
mechanics can be restated in terms of a frozen formalism. One has only to
parametrize the theory, making the time variable the $(n + 1)$st configuration
coordinate,\ldots  \cite{BergmannGeometryObservables} \end{quote} 
with $p=-H.$
The second point involves whether observables are global, local, or something else.  Observables are probably not globally defined, but probably not locally either unless by  coincidences or asymptotically. 
  Komar-Bergmann intrinsic coordinates might  make observables out of the remaining field variables.  But such an approach is no panacea. ``This approach is anything but aesthetic, or even practical, and in my opinion will serve at best only as an example guaranteeing existence.''  % my thoughts exactly.  
It seems that Bergmann's notion of observables is not flourishing even by his standards.

%%%%%%%%%%%%%%%%%%%%%%%%%%%%%%%%%%%%%%%%%%%%%%%

\subsection{Symmetries in Gauge Theories (1978)}

Bergmann and Flaherty consider the question of what symmetries ought to mean for configurations of gauge theories \cite{BergmannFlahertySymmetries}. 
They produce plausible definitions of symmetries for potentials, connections, and field strengths (recognizing that these quantities do not line up in the same way for GR as for Yang-Mills theories).  The basic notion for GR given their definitions turns out to be not Killing vector fields (vector fields  $\xi^{\alpha}$ such that $\pounds_{\xi} g_{\mu\nu}=0$), but rather the logically weaker (hence extensionally larger) notion of what Bergmann and Flaherty call affine Killing vector fields, which are also known as affine collineations, vector fields   $\xi^{\alpha}$ such that $\pounds_{\xi} \Gamma^{\alpha}_{\mu\nu}=0$ \cite{MaartensAffineCollineationRobertsonWalker}. 
While the question of observables in GR does not arise explicitly, it is noteworthy that the authors now employ for the most part and consciously \emph{active diffeomorphisms} rather than passive coordinate transformations.

%%%%%%%%%%%%%%%%%%%%%%%%%%%%%%%%%%%%%%%%%%%%%%%%%%%%%%

 \subsection{Bergmann \& Komar for Einstein's Centenary (1979/1980)}

This well known review article appeared in a work celebrating the centenary of Einstein's birth  \cite{BergmannKomar100}.  It reviews much of Bergmann's thoughts on quantization since the late 1950s.  One finds indications of an active viewpoint, such as the word ``diffeomorphism.''  General Relativity, unlike earlier theories, exhibits  ``the invariance of the laws of nature with respect to arbitrary diffeomorphic mappings of the space-time manifold on itself.''
However,  ``coordinate transformation'' and related phrases also appear very frequently.  The authors seem not to think that much is at stake in making a clear distinction.  
Part of the conceptual challenge of GR is due to the fact that choosing a coordinate(s) in one era fails to solve the problem for other eras.  ``In general relativity, a hypersurface of simultaneity no longer defines by itself a preferred congruence of spacelike hypersurfaces.''  \cite{BergmannKomar100}

One notices a friendly nod toward Hamiltonian-Lagrangian equivalence (an idea that he seems never to repudiate even when he violates it).   ``The currents $C^{\rho}$ are the generating densities of the infinitesimal coordinate transformations $\delta x^{\rho}.$ Intentionally we have introduced
them without bothering with a full-blown canonical formalism. Our $C^{\rho}$ are, however, the same generating densities defined in the canonical  context.''  \cite{BergmannKomar100}

He  and Komar continue to give Dirac credit for completing the Hamiltonian formalism.
\begin{quote}  
In 1958 P. A. M. Dirac [references suppressed] succeeded in presenting a complete Hamiltonian version of general relativity
and thereby, incidentally, demonstrating the substantive uniqueness of
propagation from Cauchy data.
 \cite{BergmannKomar100} \end{quote} If Dirac's treatment was complete (as they say), and if Dirac was explicitly willing to sacrifice the $4$-dimensional symmetry \cite{DiracHamGR}, it is no wonder that a $3+1$ gauge generator was not a priority for the later Bergmann---even if one of the main contributions of his early work with Anderson was the gauge generator \cite{AndersonBergmann}. Bergmann and Komar exhibit Dirac's Lagrangian in a form starting with the Hilbert Lagrangian, adding terms to get Einstein's Lagrangian (which removes second derivatives), and adding terms to trivialize four canonical momenta.

It is claimed that the Hamiltonian constraints  $\mathcal{H}_0$ and $\mathcal{H}_i$ generate diffeomorphisms.  
``The Hamiltonian constraints generate mappings which map equivalence classes into themselves;
that is, they merely alter the initial Cauchy surface within a given space-time.''  \cite{BergmannKomar100}
Here one does not find a clear distinction between time evolution and coordinate transformations, partly because active diffeomorphisms cloud the distinction.  
Remarks on a Hamilton-Jacobi-like formulation brings a  welcome mention of the need for phase space $\times$ time, not just phase space:  ``The product
of ordinary phase space by the time axis corresponds to the constraint
hypersurface....''

Bergmann and Komar continue to prefer  invariant (as opposed to covariant) quantities.
 \begin{quote} In a general-relativistic theory observable quantities should be
those whose values are intrinsic to a particular space-time, and independent
of the choice of coordinate system. That implies that only first-class variables
are observable. \cite{BergmannKomar100} \end{quote} 
This familiar claim appears without much novel argumentation.  
\begin{quote} One can argue that in the quantum version of a general-relativistic theory
only observables should play a role, as they are the only operators whose
action on physically permissible kets leads to physically permissible kets. In
the following sections the degree of viability of that notion will be discussed
further. It will prove desirable to introduce a modified concept, the
quasiobservables, which are the analogs of Cauchy data in the classical
formalism. \end{quote}

Quantizing using just observables does not look viable given ``that thus far no one has been able to
discover a single classical observable.''  
Indeed it seems advisable to modify the concept to a logically less restricted notion of ``quasiobservables'' in order to recover an Ehrenfest theorem, they conclude.

%%%%%%%%%%%%%%%%%%%%%%%%%%%%%%

\subsection{The Fading World Point}

This brief equation-free paper at a physics conference at Erice \cite{BergmannFadingWorldPoint} reflects Bergmann's continued interest in point identity in recognizable continuity with Einstein's point-coincidence argument and even his late remark that not even a topological space would remain if the space-time metric were removed  \cite[p. 155]{EinsteinSpaceNotLeftBehind}. 
\begin{quote} 
No matter what further structures are imposed on the space-time manifold, it
has been usually taken for granted that each world point has an
invariant identity, which is preserved under all symmetry transformations
of a given theory.
From the point of view of Poincare invariance, this is the
only sensible attitude to take. \cite{BergmannFadingWorldPoint} \end{quote} 

But matters are more complicated in General Relativity.  \begin{quote} Until a specific metric field has been introduced onto
the manifold, the construction that looks so straightforward in
conventional field theories lacks meaning. And as the metric field
is not to differ in principle from other physical fields, it looks as if the identity of a world point is inextricably bound up with
the physics, - the totality of physical fields - , present.  \cite{BergmannFadingWorldPoint} \end{quote} 
Yet in the usual way of looking at GR and unified field theories, 
\begin{quote} 
the identity of world points is preserved by the assumed symmetry group of
transformations, transformations that map one world point on another
world point without regard to the metric and other fields
supposed to exist. This symmetry group is usually referred to as
the group of diffeomorphisms (or of curvilinear coordinate transformations).  \cite{BergmannFadingWorldPoint} \end{quote}
This last passage shows again how little Bergmann was concerned to have a clear distinction between active and passive transformations.  To me this passage is not particularly clear, but later parts of the paper suggest that ``without regard to the metric and other fields'' means that the transformations do not depend on the metric or other fields.

Both Bergmann's view that Dirac's late 1950s contribution to GR was definitive and the importance that Bergmann continued to place on the Bergmann-Komar exploration of field-dependent transformations become evident.
\begin{quote}
Dirac finally constructed a commutator in three dimensions
that did not require off-surface derivatives, and thereby succeeded
in completing the Hamiltonian formulation of general relativity [endnote suppressed].
His device, formally speaking, consisted of replacing partial 
derivatives of the metric along the $x^0$-axis (``time'' derivatives) by derivatives in the direction perpendicular to the chosen Cauchy
hypersurface. The generators of these infinitesimal mappings, no
longer ``pure'' diffeomorphisms (i.e. mappings irrespective of any
metric) were the so-called Hamiltonian constraints,\ldots  \cite{BergmannFadingWorldPoint}
\end{quote}

Bergmann draws a conclusion and suggests its possible future utility.    
\begin{quote} What this whole analysis may teach us is that the world point
by itself possesses no physical reality. It acquires reality only
to the extent that it becomes the bearer of specified properties
of the physical fields imposed on the space-time manifold.   \cite{BergmannFadingWorldPoint}
\end{quote} 
 The essay  shows various continuing tendencies in Bergmann's thought.

%%%%%%%%%%%%%%%%%%%%%%%%%%%%%%%%%%

\subsection{Bergmann 1986/1987/1989 on Canonical GR, 1930-1959}

This paper on the history of canonical quantization gives Bergmann's late views on the lasting achievements from the first three decades of work, which included from 1949-1959 the reinvention of much of what came before, making the 1950s the key decade \cite{BergmannEarly}. (Almost the same work was also published elsewhere  \cite{BergmannCharapTetrad}.)  One of the key themes is how highly Bergmann viewed Dirac's contributions to simplifying the formalism in ways that diminished manifest $4$-dimensional space-time symmetry.  

The claim that first-class constraints (separately) generate gauge transformations is evident in Bergmann's preparation for a discussion of reduced phase space.
\begin{quote} 
If the infinitesimal group of mappings that is generated by the first-class
constraints is expanded into a group, a group of canonical mappings that
map the constraint hypersurface on itself, then those orbits of that group of
mappings that have at least one point on the constraint hypersurface lie in
that hypersurface entirely; we call them \emph{equivalence classes}. These equivalence
classes cover the constraint hypersurface, and do so without overlap. Each
point on the constraint hypersurface belongs to exactly one equivalence class.

Equivalence classes correspond to sets of values of the canonical field
variables that are carried over into each other by members of the symmetry
group of the theory. In the language of physics, we should say that the points
of one equivalence class represent the variety of ways how one-and-the-same
physical situation can be represented. Distinct equivalence classes correspond
to inequivalent physical situations. 
\cite{BergmannEarly} 
\end{quote}

This belief that constraints separately generate gauge transformations leads to the expected conclusions about observables.
\begin{quote} 
Dynamical variables that are defined on the reduced phase space are those
that are constant over each equivalence class, and hence, by definition, invariant
with respect to the symmetry group of the theory. They are \emph{observables},
in the sense that their values depend only on the physical situation, not on
the manner in which we present it. Poisson brackets between observables are
well defined, though not between other dynamical variables.

All of these facts were worked out in great detail during the 1950s. \cite{BergmannEarly}
\end{quote} 
Clearly Bergmann remained satisfied with these ideas, in contrast (as we have seen) to the changes in his view of the value of defining local observables by coincidences.

Belief in gauge generation by first-class constraints apparently remains psychologically compatible with belief in the gauge generator, because the next sentence in the same paragraph recalls work (which one finds earlier in the decade \cite{AndersonBergmann}) about the gauge generator.
\begin{quote} 
Fairly early in that decade it was also established that the role of the primary,
secondary, etc., constraints had to do with the form of the (infinitesimal)
transformation laws for the field variables. Such transformation laws might
contain references to the characteristic functions of a coordinate or gauge
transformation (such as the vector field representing an infinitesimal coordinate
transformation), to their time derivatives, to their second time derivatives,
and so forth. Primary constraints must be multiplied by the highest-order time
derivatives, secondary constraints by the next-highest-order time derivatives,
and so on, to yield the appropriate generators of the whole transformation. \cite{BergmannEarly} 
\end{quote}  
Since the mid-1950s constrained Hamiltonian dynamics has been a field in which new ideas are introduced alongside previous ideas that they contradict, but this contradiction is generally not noticed.  This fact plays a key part in setting up the so-called ``problem of time.''

In section 2 one finds Bergmann thinking in terms of active diffeomorphisms, a practice that was quite absent from the first two decades or more of his work on canonical GR.
\begin{quote} 
The infinitesimal mapping that is supposed to contain the dynamical essence
of a physical theory is the displacement in time, that is to say the space-time
mapping $\delta t = a.$ This mapping certainly is a diffeomorphism; it belongs to
the symmetry group of any general-relativistic theory. 
\end{quote} 
This shift is somewhat surprising considering how often the problem of point individuation and the  need for a relational solution to it (such as in terms of coincidence observables using Weyl scalars) appeared in his earlier work.  
The shift seems to have bled into his memory of his and Dirac's work.  ``In the early 1950s, both Dirac and we made an attempt to disconnect the
group of diffeomorphisms from the dynamics of the theory by introducing a subsidiary set of coordinates, which we called \emph{parameters}.''

Bergmann is perhaps overly generous towards Dirac's contribution  regarding the rejection of parameters. 
\begin{quote} 
Subsequently, Dirac found that
the introduction of parameters is not necessary for the construction of a
Hamiltonian formalism. Without them the whole formalism is much less
involved. As it turns out, in Dirac's formulation the primary constraints can
be eliminated, and the Hamiltonian density is a linear combination of first-class
secondary constraints, the coefficients being arbitrary functions of the
canonical field variables and the coordinates. \cite{BergmannEarly} 
\end{quote} 
But Bergmann's group already found it expedient to reject parameters years earlier \cite{PenfieldNoParameters,AndersonBergmann,GoldbergSyracuse,SalisburyBergmann,BlumSalisburyRickles,SalisburySyracuse1949to1962}. Something that  Dirac \emph{did} accomplish in 1958 of undeniable lasting value (though paralleled by James L. Anderson's work \cite{AndersonPrimary} in a more canonical form) is, perhaps among other things, to find a total divergence to add to the Lagrangian so that the canonical momenta were nicely changed from a collection of $10$ of which only $6$ were independent in the usual way, to a collection of $6$ which were independent and conveniently related to the spatial metric and a collection of $4$ which simply vanish, thus radically simplifying the primary constraints.  Further  conclusions apparently drawn by both Dirac \cite{DiracHamGR} (quite explicitly) and Bergmann include the possibility of dropping those $4$ canonical momenta from the phase space (which implies that the Hamiltonian does not even formally determine the evolution of $g^{00}$ and $g_{0r}$) and that retaining any explicit relationship to $4$-dimensional tensor calculus was unnecessary and counterproductive.  Hence neither space-time covariance nor mathematical $H$-$L$ equivalence was a priority---even beyond the slight to those ideas already implied by the idea that first-class constraints \emph{separately} generate gauge transformations.  
Such views would explain why there seems to have been no effort to reinvent the Rosenfeld-Anderson-Bergmann gauge generator in $3+1$ form from the Syracuse group, a problem resolved in the 1980s \cite{CastellaniGaugeGenerator}.

The latter parts of this chapter also show how decisive Bergmann took Dirac's late 1950s contributions to have been.
One problem is that trying to represent temporal coordinate transformations seemed to lead to ever higher time derivatives, beyond what the canonical formalism could accommodate. (Perhaps this was seen as a flaw in the Anderson-Bergmann gauge generator? See also (\cite{SalisburySyracuse1949to1962}.)  
\begin{quote} 
 Thus the generator of the commutator must contain elements that cannot possibly be obtained from a
canonical commutation of whatever ingredients are available on the Cauchy
surface! An adequate description of the symmetry group of general relativity
in terms of a canonical formalism appears foreclosed!

This difficulty was overcome by Dirac in his 1958 paper.\ldots

Dirac's procedure at first appeared like magic.\ldots  

Aside from overcoming the seemingly insurmountable obstacle presented by
the peculiarities of the group of(four-)diffeomorphisms, Dirac streamlined the
identification of the canonical variables. In the earlier formulations there had
been primary and secondary constraints. Almost simultaneously, DeWitt and
Anderson had discovered a canonical transformation that simplified the primary
constraints, and so had Dirac. By making the primary constraints purely
algebraic, Dirac was able to eliminate from the formalism four pairs of
canonically conjugate variables, $g_{0\mu}$ and $\pi^{0\mu}$, retaining only the three-metric
$g_{mn}$ and their canonically conjugate momentum densities.\ldots

Dirac essentially completed the canonical formulation of general relativity.
His procedure would presumably also be applicable to modified theories of
gravitation, such as the scalar-tensor theories and the Einstein-Cartan theories,
provided there is a metric structure. \cite{BergmannEarly}
 \end{quote} 
That Dirac's streamlining implies the \emph{de facto} elimination of $4$-dimensional coordinate symmetry and makes the evolution of $g_{0\mu}$ inexpressible even formally from the Hamiltonian is not something that Bergmann emphasizes.

%%%%%%%%%%%%%%%%%%%%%%%%%%%%%%%%%%%

\subsection{Observables in General Relativity (1988)} 

This not well-known conference paper synthesizes Bergmann's mature  thinking about observables \cite{BergmannObservablesGravitationalMeasurements}, a topic on which he had written for over three decades.  This 4-page work summarizes a great many of the things that he had previously said, including an expectation that observables are closely tied to observations and  a claim that observables should Poisson-commute with Dirac's Hamiltonian constraints (note: not the gauge generator and not the primary constraints along with Dirac's $\mathcal{H}_0$ and $\mathcal{H}_i$).  He infers from the  Poisson bracket claim that observables do not change, a conclusion that he rightly finds awkward and in tension with the connection to observations (the latter a topic to which he returns at the end of the paper).   Again he does not make the link to the transport term in the Lie derivative to recognize the absurdity of demanding invariance under a displacement as a condition of gauge invariance, however \cite{GRChangeNoKilling}.

The Komar-Bergmann intrinsic coordinate scheme in terms of Weyl scalars is recalled; while it can convert components of tensors into observables, it is ``rather unwieldly'' \cite{BergmannObservablesGravitationalMeasurements}. More specifically, this approach does not work in the presence of metric symmetries and is very remote from quasi-Lorentzian coordinates typically employed. Maximal surfaces, arbitrary hypersurfaces pierced by a congruence of geodesics, and other ways of fixing the coordinate system are contemplated.  One might wonder, however, whether an effort to elevate some coordinates over others, especially in a quantitatively precise way (as opposed, say, to preferring coordinates that are more or less lengths, perhaps those admitting a series expansion  \cite{BorisovOgievetskii} in terms of other coordinates that are more or less lengths, over coordinates that are angles) wouldn't subvert one of the main features of GR.

 Bergmann  concludes that the Poisson bracket of observables should be $0$ based largely on analogy with electromagnetism or gauge theories in the narrow sense.
\begin{quote} 
 Just as an instrument cannot take into account our choice of a gauge frame, it cannot be programmed to act within a definite (curvilinear)
coordinate system, as these choices are made within our minds, or on a piece of paper, not by interaction with the physical universe.
\end{quote} 
This passage and others show Bergmann's continuing link between observables and observations; this one also motivates the invariance of observables under internal gauge transformations.   Coordinates, however, are things to which we can point in the sense of saying that the here and now (as I type this, say) has certain coordinate values; recall the golf course on the Prime Meridian near Cambridge.  
Hence observation-relative-to-a-coordinate system, with the coordinates themselves associated with features of the world, makes more sense than Bergmann fears. Traditionally one took the infinitesimal interval $ds^2 = g_{\mu\nu} d^{\mu} dx^{\nu}$ to be observable in the ordinary sense.  The idea was not that the world supplied the infinitesimal displacement and coordinate expression thereof $dx^{\mu},$ but that one could point to various pairs of  nearby space-time events, give their coordinate differences in any convenient coordinate system, and thereby synthesize the space-time metric relative to that coordinate system;  one could also transform the result to another coordinate system and could  sew together results from multiple coordinate systems using the tensor transformation law.  Hence the metric was observable relative to a coordinate system:  covariant, not invariant. I do not see why this picture or something like it is inadequate.  Indeed the striking empirical success of GR, which must involve processes sufficiently similar to this sketch, indicates that in general outline it is correct. The many GR textbooks that explain the theory, its empirical predictions, and something of those predictions' success, without mentioning Bergmann's concept of observables, bear this out.    Hence his longstanding quest for ``observables'' that are predictable full-stop, rather than predictable-up-to-coordinate-transformations, strikes me as simply unmotivated in classical GR, a highly empirically successful theory. The problem of observables classically is a pseudo-problem, the answer being just $4$-dimensional tensor calculus all over again (with Legendre transformations if a Hamiltonian formalism is used) \cite{ObservablesEquivalentCQG}.  Predictability full-stop or having $0$ Poisson brackets might  be useful on technical grounds for quantization, but those are quite different arguments with little direct connection to everyday observations.

%%%%%%%%%%%%%%%%%%%%%%%%%%%%%

\subsection{Quantization of the Gravitational Field, 1930-1988 (1988/1992)}

This very brief (3-page) paper does not actually address the subject of observables at all 
 \cite{BergmannQGF}, but it  give further confirmation of the later Bergmann view that Dirac had made a decisive contribution in 1958 in the work that trivialized the primary constraints and beyond.
\begin{quote} 
After World War II, Paul A.M. Dirac (Dirac 1950,
1951; Pirani and Schild 1950) and those of us at Syracuse (Bergmann
et al. 1950) initiated the canonical formulation of general relativity as an
initial step toward quantum gravity. These efforts culminated in Dirac's
hands in complete success (Dirac 1958, 1959). The Lie algebra initiated
by Dirac was closed more than 10 years later (Bergmann and Komar
1972).
\end{quote} 
There is no doubt that earlier versions of canonical GR, such as those in which the $4$ primary constraints tied together the $10$ canonical momenta because of a matrix of rank $6$    \cite{RosenfeldQG,PiraniSchild,BergmannPSZ,DeWittSpinor,BelinfanteCK} 
were  difficult to use and difficult to understand.  It is also clear that trivializing the primary constraints (something achieved almost simultaneously by Anderson \cite{AndersonPrimary} and B. DeWitt (unpublished) by somewhat different means), was a great technical advance that also was highly suggestive in matters of interpretation.  The question is whether such suggestions should have been accepted in the strongest possible way without adequately weighing the cost.  Consider that one can add a divergence to Maxwell's electromagnetism to include the expression $(\partial_{\mu} A^{\mu})^2$ in the Lagrangian (now gauge quasi-invariant instead of invariant)  and thereby de-trivialize the primary constraint.  While this is a step in the wrong direction, its possibility makes it perhaps somewhat less surprising that there might exist a divergence to add to GR to trivialize the primaries. It is in any case difficult to see why adding a divergence to the Lagrangian, something that one always knew to be possible in general, should imply also that the phase space should be shrunk and that space-time symmetry should be radically obscured or discarded.  Such is one possible interpretive conclusion, but by no means the only one, and is quite radical.  What seems to have been missing prior to the 1980s  \cite{CastellaniGaugeGenerator,SalisburySundermeyerEinstein} was a  Goldilocks approach that was not so $4$-dimensional as to reproduce the mess of the early 1950s and not so $3$-dimensional as to lose the ability to express $4$-dimensional coordinate transformations in a canonical formalism.   It is probably not coincidental that Bergmann, who often emphasized GR-exceptionalist views among the foundations of his nonperturbative canonical approach to quantum gravity, would be predisposed to accept more revolutionary novelty than was required and that someone with more connections to particle physics (Castellani) would engage in nuts-and-bolts calculation to find the $3+1$ gauge generator and thus find a middle path.  While one thinks of Dirac as a particle physicist in his early work, his emphasis on aesthetic criteria in physics clearly points away from the non-philosophical nuts-and-bolts calculation stereotype in favor of mental habits more akin to those of general relativists.  Particle physics can be useful for conceptual problems in GR \cite{TenerifeProgressGravity,LambdaMPIWG}.

%%%%%%%%%%%%%%%%%%%%%%%%%%%%%%%%%%%%%%%%%%%%

\subsection{Bergmann at Erice, 1991}

The last Bergmann work to be considered is a very interesting and wide-ranging survey of GR with an eye to its relevance to astronomy, including conservation laws.  The portions bearing on the Hamiltonian treatment of GR hold few surprises.  The role of the true degrees of freedom in electromagnetism as patterns to be emulated in GR is evident.   ``$\phi$ is not a dynamical variable, and neither is the longitudinal
part of the $\vec{A}$-field.''  Dynamical variables here are used in a rather restricted sense, evidently.   ``In general relativity the situation is similar.'' However,  ``\ldots there are no simple `gauge-invariant' field
quantities.''  The issue of space-time point individuation remains of interest.  ``How can we identify a point independently of the fields there? Certainly not by the numerical values of its coordinates!''

The Hamiltonian formalism leads to various difficulties. 
 \begin{quote}  All attempts at quantization give rise to a host of
additional technical and conceptual problems.\ldots % not all of which have been resolved.

I shall mention but a few. As the constraints are the
generators of propagation from one hypersurface to another, the
 ``observables'', the variables that correspond to gauge-invariant
field variables in electrodynamics, must commute (or have vanishing
Poisson brackets) with the constraints, which is to say, they must
be constants of the motion.
\cite{BergmannCornerstone}
\end{quote}  
One notices both the use of first-class constraints separately and the quest for invariant (as opposed to covariant) quantities, the two aspects of the definition of observables that merit rethinking, as well as the electromagnetic precedent.  
\begin{quote} %\ldots 
 The generators of coordinate transformations are known, they are the
``constraints'', the expressions of the canonical variables that must
vanish on every Cauchy hypersurface \underline{a priori}, before they are used
to propagate to neighboring hypersurfaces. They are ``first class''\ldots  \cite{BergmannCornerstone} \end{quote} 
Constraints are used separately again.  
Yet there is also much talk of  $4$-dimensional coordinate transformations without noticing the tension.   One can say, at any rate, that Bergmann displayed a quite stable mix of views for  nearly 40 years, regardless of their logical compatibility.  
This stability is also reflected in J. Goldberg's history   \cite{GoldbergSyracuse}.  %  Affirms the conventional wisdom in defining observables, etc., describes as key results

%%%%%%%%%%%%%%%%

 \section{Conclusions about Bergmann and Observables}

In surveying more or less everything that Bergmann wrote on observables, one encounters many interesting and plausible ideas and many repeated themes, but  no  coherent view. There arose a tendency to lose  the $4$-dimensional symmetry partly due to neglecting $G$ in favor of separate first-class constraints. Requiring a vanishing Poisson bracket with each first-class constraint is logically stronger than requiring vanishing Poisson bracket with $G$, so that could by itself resort in a shortage of observables.    Failure to reckon with the external \emph{vs.} internal distinction (or something in the neighborhood) led to an unmotivated requirement of a $0$ Poisson bracket.  A $0$ Poisson bracket implies not mere invariance in the sense of a scalar function, but vanishing Lie derivative and hence invariance even under the transport term.  Such a requirement is the infinitesimal analog 
of requiring sameness between 1 a.m. daylight savings time and 1 a.m. standard time.  Bergmann  understood the classical Lie derivative perfectly well, but somehow imposed such a requirement anyway, an enduring puzzle.  On various occasions Bergmann endorsed  $H$-$L$ equivalence for observables and  never rejected it, even while also holding views that contradict it.  At times he wanted local observables; sometimes he claimed to have them using intrinsic coordinates. But he also inferred from this technical Poisson bracket condition that observables do not change.  Bergmann tried to attach too many  nice features to  a single concept and did not arrive at a logically consistent view.  

  The  desiderata of local observables and $H$-$L$ equivalence are fulfilled when one takes observables to be space-time tensors or geometric objects more generally (assuming also internal gauge invariance as appropriate), as the beginning of the paper argued.  Reflecting on the classical derivation of the Lie derivative and on  equivalent observables for equivalent theories gives largely independent arguments for defining observables $\{O, G \} \sim \pounds_{\xi} O $.  Observables are invariant under internal transformations, but covariant under coordinate transformations.  Given that some quantization programs have been based around observables, reflection on the nature of observables might shed some light on quantum gravity research programs. On the view of observables defended here, observables are fairly common and familiar, but possibly less suited to quantization than one might have hoped.  Perhaps such  quantization programs are proceeding  perfectly reasonably, but are using technical concepts that should not be called observables.  

% 
%
%
%

%%%%%%%%%%%%%%%%%%%%%%%%%%%%%%%%%%%%%%%%%%%%

\section{Appendix: Dirac and Observables}

While the paper is about the work of Bergmann, Dirac is the other major figure working on the same topic for several decades, so his work merits attention. 
Dirac does not mention the term ``observables'' in his book \cite{DiracLQM} or in other papers that I have checked recently, but his remarks about leaving the \emph{state} the same while changing the $q$'s and $p$'s (pp. 19-21, 28, 29) and his claim to have shown that a first-class constraint (by itself) generates a gauge transformation, however, indicate what he is likely to have thought.
There is  some basis for the widely used term ``Dirac observables'' as quantities with vanishing Poisson brackets with each first-class constraint---an idea that Bergmann sometimes endorsed, as seen above;  if one needs to ascribe a view on observables to Dirac, then the usual ascription implied by the term ``Dirac observables'' is plausible (unlike Bergmann and ``Bergmann observables'').  Dirac's  association of states with values of $t$ in this early chapter of the book makes the ideas ill-adapted to General Relativity with its velocity-dependent gauge transformations, while the later chapters working with curved surfaces do not engage as clearly with quantities that are invariant under what are taken to be gauge transformations.

He came a bit closer in accommodating General Relativity in  earlier work. 
\begin{quote} 
Different solutions of the equations of motion,
obtained by different choices of the arbitrary functions of the time with given initial
conditions, should be looked upon as all corresponding to the same physical state
of motion, described in various way by different choices of some mathematical
variables that are not of physical significance (e.g. by different choices of the gauge
in electrodynamics or of the co-ordinate system in a relativistic theory). \cite{DiracHam}
\end{quote} 
Instead of the seemingly precise (if partly misdirected) notion of an instantaneous state of the world at a given time, here Dirac uses an ambiguous term ``state of motion,'' not clearly distinguishing  states from histories.  
A paper devoted to General Relativity gives attention to instantaneous states and suggests that the simplification of Hamiltonian methods suggests giving up four-dimensional symmetry \cite{DiracHamGR}.  A later conference paper on quantum gravity was devoted primarily to technical rather than conceptual questions \cite{DiracQGConstraints}. His posthumously published 1975 Florida State lectures on General Relativity do not address Hamiltonian methods \cite{DiracGR}.

One also recalls that  Dirac's argument that a first-class constraint generates a gauge transformation, which is followed in detail by Henneaux and Teitelboim \cite{HenneauxTeitelboim}, is fallacious \cite{PonsDirac,FirstClassNotGaugeEM}.  One problem is that in equation 1.37, Dirac has temporarily forgotten that his expression $H^{\prime}$ for electromagnetism contains a term involving the secondary first-class constraint (the phase space Gauss law condition) multiplied by the gauge-dependent quantity $- A_0$.  Consequently he thinks that his quantity $v_a$ is completely arbitrary, when in fact it is closely tied (through time differentiation) to the coefficient of a first-class secondary constraint inside $H^{\prime}$---as he himself notices later on p. 26.  The question is not what did Dirac know, or what did he assert in chapter 2, but what did he 
\emph{argue} in chapter 1.  While chapter 2 \emph{illustrates} his ideas in application to electromagnetism, the example is not allowed to push back on the earlier material.  There is no (successful) quest for  reflective equilibrium between general principles and independently understood examples.  Hence electromagnetism illustrates not the separate gauge-generating activity of each first-class constraint, but the coordinated team activity of the primary and secondary first-class constraints together.

%%%%%%%%%%%%%%%%%%%%%%%%%%%%%%%%%%%%%%%%%%%%

\section{Acknowledgments}

Thanks to Alex Blum for polishing translation of Bergmann's  Bern paper excerpts and for discussion  at talks in Berlin, Jeremy Butterfield for helpful remarks on exposition and Frege over the years, and Don Salisbury for bibliographic  help in finding relevant Bergmann works, help seeing Schiller's thesis, and years of discussion.

%\bibliography{Pitts}  
%%%
%\bibliographystyle{apalike} 
%%
%%
\end{document}